\documentclass[]{pasj02}
\usepackage[switch,mathlines]{lineno}
\usepackage{natbib}
\usepackage{wrapfig}
\usepackage{url}
\usepackage[normalem]{ulem} 

\jyear{2026}
\Received{2026 February 3}
\Accepted{2026 March 30}

\begin{document}

\title{
A XRISM Study of Highly Ionized Iron Emission Lines \\ from the Low-Eddington-ratio AGN in NGC\,7213}

\author{
\raggedright
Kaito~\textsc{Murakami},\altaffilmark{1}\altemailmark\email{murakami@ess.sci.osaka-u.ac.jp}
Taiki~\textsc{Kawamuro},\altaffilmark{1}\altemailmark\email{kawamuro@ess.sci.osaka-u.ac.jp}\orcid{0000-0002-6808-2052}
Ryota~\textsc{Tomaru},\altaffilmark{1}\altemailmark\email{r.tomaru.sci@osaka-u.ac.jp}\orcid{0000-0002-6797-2539}
Hirokazu~\textsc{Odaka},\altaffilmark{1}\orcid{0000-0003-2670-6936}
Elias~\textsc{Kammoun},\altaffilmark{2}\orcid{0000-0002-0273-218X}
Shoji~\textsc{Ogawa},\altaffilmark{3}\orcid{0000-0002-5701-0811}
Stefano~\textsc{Bianchi},\altaffilmark{4}\orcid{0000-0002-4622-4240}
Hirofumi~\textsc{Noda},\altaffilmark{5}\orcid{0000-0001-6020-517X}
Claudio~\textsc{Ricci},\altaffilmark{6,7}\orcid{0000-0001-5231-2645}
Yuichi~\textsc{Terashima},\altaffilmark{8}\orcid{0000-0003-1780-5481}
Yoshihiro~\textsc{Ueda},\altaffilmark{9}\orcid{0000-0001-7821-6715}
Satoshi~\textsc{Yamada},\altaffilmark{5,6,10}\orcid{0000-0002-9754-3081}
and
Hironori~\textsc{Matsumoto}\altaffilmark{1}
}

\altaffiltext{1}{Department of Earth and Space Science, Graduate School of Science, The University of Osaka, 1-1 Machikaneyama, Toyonaka 560-0043, Osaka, Japan}
\altaffiltext{2}{Cahill Center for Astronomy \& Astrophysics, California Institute of Technology, 1216 East California Boulevard, Pasadena, CA 91125, USA}
\altaffiltext{3}{Japan Aerospace Exploration Agency (JAXA), Institute of Space and Astronautical Science (ISAS), Chuo-ku, Sagamihara, Kanagawa 252-5210, Japan}
\altaffiltext{4}{Dipartimento di Matematica e Fisica, Universit\`a degli Studi Roma Tre, via della Vasca Navale 84, I-00146 Roma, Italy}
\altaffiltext{5}{Astronomical Institute, Tohoku University, Miyagi 980-8578, Japan}
\altaffiltext{6}{Department of Astronomy, University of Geneva, ch. d’Ecogia 16, 1290, Versoix, Switzerland}
\altaffiltext{7}{Instituto de Estudios Astrof\'isicos, Facultad de Ingenier\'ia y
Ciencias, Universidad Diego Portales, Av. Ej\'ercito Libertador 441, Santiago, Chile}
\altaffiltext{8}{Department of Physics, Ehime University, Ehime 790-8577, Japan}
\altaffiltext{9}{Department of Astronomy, Kyoto University, Kyoto 606-8502, Japan}
\altaffiltext{10}{Frontier Research Institute for Interdisciplinary Sciences, Tohoku University, Sendai 980-8578, Japan}


\maketitle

\begin{abstract}
We present an analysis of XRISM and NuSTAR data obtained for the nearby low-Eddington active galactic nucleus NGC\,7213. 
Our goal is to examine whether its He-like and H-like iron emission lines can be reproduced by photoionization or collisional ionization processes. 
Using the broad-band energy coverage of our data (2--60~keV), we first constrained the continuum shape. Then, we focused on the iron-K band in the Resolve spectrum. 
Gaussian fits to Fe {\sc xxv} He$\alpha$ and Fe {\sc xxvi} Ly$\alpha$ lines suggest that they may have different velocity widths: $v_\sigma=790^{+370}_{-240}$~km~s$^{-1}$ for Fe {\sc xxv} and $v_\sigma=2610^{+1700}_{-1580}$~km~s$^{-1}$ for Fe {\sc xxvi}. 
In this case, the He$\alpha$ resonance line (w) and forbidden line (z) have similar intensities of $\approx0.5$--$0.6\times10^{-5}$~ph~s$^{-1}$~cm$^{-2}$, while 
the intercombination lines (x+y) are not significantly detected with upper limits of $\lesssim 0.2\times10^{-5}$~ph~s$^{-1}$~cm$^{-2}$. 
Motivated by the possible difference in the line widths, we tested one- and two-zone photoionized and collisionally ionized models. 
Our results show that the additional ionized component is not significantly required, and the current data cannot uniquely determine whether photoionization or collisional ionization dominates. 
Moreover, if the Fe\,{\sc xxv} He$\alpha$ complex implies that the weak x+y lines are suppressed relative to the  w and z lines, such a structure is difficult to reproduce with either ionization model adopted. 
Finally, by comparing NGC\,7213 with M\,81$^\ast$, accreting at a much lower Eddington ratio of $\lambda_{\rm Edd}\sim 10^{-5}$, we found a decrease in the density of the gas responsible for highly ionized iron emission, which may imply that the density decreases with decreasing $\lambda_{\rm Edd}$.

\end{abstract}

\section{Introduction}
Active galactic nuclei (AGNs) are powered by mass accretion onto supermassive black holes (SMBHs) and are composed of multiple structural components, including an accretion disk, a hot X-ray emitting corona, broad- and narrow-line regions, and outflows (e.g., \citealt{Peterson1997,Netzer2015,RamosAlmeida2017}).
These structures are thought to change with the Eddington ratio, $\lambda_{\rm Edd}$ \citep{Ho2008,Elitzur2009,Trakhtenbrot2017}.
At high Eddington ratios ($\lambda_{\rm Edd}\gtrsim0.1$), a geometrically thin, optically thick accretion disk is expected to extend down to an inner orbit and is often accompanied by radiation-driven winds (e.g., \citealt{King2003,Takeuchi2013,Giustini2019,Nomura2020}).
In contrast, in low-luminosity AGNs (LLAGNs) accreting at $\lambda_{\rm Edd}\lesssim10^{-2}$, the inner disk may be truncated at larger radii and replaced by a geometrically thick, optically thin, radiatively inefficient accretion flow (RIAF; e.g., \citealt{Narayan1994,Yuan2014}).
In this regime, the driving mechanism of outflows may be  different; thermal buoyancy and magnetic forces may work (e.g., \citealt{Yuan2014}).

X-ray spectroscopy of highly ionized iron features has been an invaluable way to probe the physical conditions and origin of gas in the nuclear region, including both the accretion flow and outflows.
He-like Fe {\sc xxv} and H-like Fe {\sc xxvi} emission and absorption lines around 6.7 and 6.97~keV arise in highly ionized gas with temperatures above $\sim10^{7\text{--}8}$~K and/or under strong photoionizing radiation. Thus, they can trace material located close to the mass-accreting SMBH that is difficult to study at other wavelengths (e.g., \citealt{Ross1999,Nayakshin2001,Kallman2001,Bianchi2002}).
From the centroid energies and widths of these lines, we can measure the bulk velocity and velocity width of the gas, while the relative strengths of Fe {\sc xxv} and Fe {\sc xxvi} and, when resolved, the fine-structure line ratios are capable of constraining its ionization state, column density, and temperature.
In several nearby well-studied LLAGNs, highly ionized iron emission has been detected and often interpreted as arising from hot gas associated with a RIAF and/or a hot wind launched from it (e.g., \citealt{Young2007,Shi2022}).
In luminous AGNs, highly ionized iron features have been widely used to study ultra-fast outflows (UFOs) with velocities of $\sim0.1$--$0.3c$ ($c$ is the speed of light; e.g., \citealt{Tombesi2010,Gofford2015}) and also the ionization state of the inner accretion disk \citep[e.g.,][]{Ballantyne2011}.  

The X-ray Imaging and Spectroscopy Mission (XRISM) \citep{Tashiro2025} now provides an opportunity to systematically investigate these iron lines with unprecedented spectral resolution and collecting area in the Fe-K band.
Its microcalorimeter spectrometer Resolve achieves an energy resolution of $\sim 5$~eV at 6~keV, allowing the Fe {\sc xxv} He$\alpha$ and Fe {\sc xxvi} Ly$\alpha$ lines to be resolved and their kinematics to be measured with high precision. 
XRISM observations of AGNs in the performance verification (PV) phase have already demonstrated the scientific potential of this capability.
In the luminous quasar PDS\,456 and the ULIRG IRAS\,05189$-$2524, XRISM revealed multiple clumpy UFO components with velocities $v_{\rm out}\simeq0.1$--$0.3c$ \citep{xrism2025PDS456,Noda2025}.
In NGC\,3783, XRISM captured the launch of a UFO associated with a soft X-ray/UV flare \citep{Gu2025}.
For LLAGNs, Resolve spectroscopy of M\,81$^\ast$ revealed precise profiles of Fe {\sc xxv} and Fe {\sc xxvi} emission lines and provided an opportunity to discuss the origin of highly ionized plasma in greater detail than ever before \citep{Miller2025}.
Although these early XRISM results already cover a wide range of Eddington ratios, there remains a large gap in Eddington ratio between M\,81$^\ast$ and the other AGNs. 
To better understand how the accretion flow and winds evolve across the Eddington ratio, it is important to observe AGNs with intermediate Eddington ratios. 

With XRISM and NuSTAR, we observed NGC\,7213 ($z = 0.005839$), one of the best galaxies hosting an LLAGN for inclusion in the XRISM sample to discuss the dependence of AGN structure on the Eddington ratio. 
The Eddington ratio is $\sim 10^{-3}$, and its short disLLance of 22.75 Mpc\footnote{The distance is estimated with the redshift of $z$ = 0.005147 with respect to the cosmic microwave background (NASA/IPAC Extragalactic Database) and the cosmological parameters of $H_{0} = 67.8~$~km~s$^{-1}$~Mpc$^{-1}$, $\Omega_{\mathrm{M}} = 0.308$, and $\Omega_{\Lambda} = 0.692$.} makes it possible to obtain good quality X-ray spectra. 
The nucleus of this source exhibits broad H$\alpha$ emission, characteristic of Seyfert 1 galaxies \citep{Phillips1979}. 
The black hole mass is estimated to be $\sim$$10^{8} M_{\odot}$ from stellar velocity dispersion \citep{Woo2002}; thus, the Eddington luminosity is $L_{\rm Edd} \approx 10^{46}$~erg~s$^{-1}$. 
Although the 2--10~keV luminosity shows variability by a factor of a few \citep{Yan2018}, \cite{Bianchi2008} reported a value of $L_{2-10\ \mathrm{keV}}\simeq1.7\times10^{42}$~erg~s$^{-1}$. 
Using a bolometric correction factor of $f_X = 10$ \citep{Duras2020}, the bolometric luminosity is estimated to be $L_{\mathrm{bol}}\simeq1.7\times10^{43}$~erg~s$^{-1}$, which corresponds to an Eddington ratio of $\lambda_{\mathrm{Edd}}\simeq1.3\times10^{-3}$. 
In previous observations of NGC\,7213, blue-shifted Fe He$\alpha$ and Fe Ly$\alpha$ lines were reported. 
The origin of these lines is discussed in terms of either photoionization or collisional ionization processes \citep[e.g.,][]{Bianchi2008,Lobban2010}.

Our first XRISM study of NGC\,7213 outlined properties of iron emission lines \citep{Kammoun2025}.
Therein, we showed that the neutral Fe K$\alpha$ emission was well described by a narrow component with a full width at half maximum 
(FWHM) of $= 650^{+240}_{-220}$~km~s$^{-1}$ and a broad component with an FWHM of $= 4500^{+5500}_{-2200}$~km~s$^{-1}$, with only a weak contribution from the reflection continuum by neutral material. 
We also found that the velocity widths of the Fe {\sc xxv} He$\alpha$ line (FWHM $=2350^{+1600}_{-900}$~km~s$^{-1}$) and the Fe {\sc xxvi} Ly$\alpha$ line (FWHM $=6100^{+2500}_{-2300}$~km~s$^{-1}$) may be different. 

In this paper, we investigate in greater detail the highly ionized iron lines.
To reveal their emission mechanisms and origins, particularly photoionization modeling was devised so that we could survey large parameter spaces efficiently (\S\ref{ssec:32}).  
The paper is structured as follows. 
\S\ref{sec:obs_datred} describes the XRISM and NuSTAR observations and their data reduction.  
\S\ref{sec:spec_ana} presents our spectral analysis to examine whether either photoionization or collisional ionization can adequately explain the highly ionized iron lines. 
In \S\ref{sec:4}, we discuss the results derived from the spectral analysis. 
Finally, \S\ref{sec:conc} provides conclusions of this work.

\section{Observations and data reduction}\label{sec:obs_datred}
We observed NGC\,7213 with XRISM, carrying two detectors of Resolve \citep{Ishisaki2025} and Xtend \citep{Noda2025}, during 2024 November 4‒7 (Observation ID: 201115010). 
Resolve was operated with an open filter while the gate valve was closed, and Xtend was operated in the 1$/$8 window mode. 
NuSTAR \citep{Harrison2013} also observed the target simultaneously with  XRISM during  2024 November 5‒6 (Observation ID: 91001636002). 
Throughout the NuSTAR observation, no serious background flare was seen.

The XRISM and NuSTAR data were reprocessed and reduced essentially in the same way as adopted in the first paper of \citet{Kammoun2025}; we thus explain our reprocessing and reduction techniques briefly. 
We reduced cleaned XRISM data provided by the XRISM team using the HEASoft 6.35.1 software and an appropriate calibration database (CALDB version 20250315), following the standard data-screening procedure\footnote{https://heasarc.gsfc.nasa.gov/docs/xrism/analysis/abc\_guide/xrism\_abc.pdf}. 
For Resolve, we used only high-resolution primary events from all 36 pixels, excluding pixels 12 and 27. 
The exclusion is because pixel 12 was used for calibration, and 
pixel 27 showed random gain jumps and its calibration was not established.
A non-X-ray background (NXB) spectrum was generated by using the \texttt{rslnxbgen} tool, whereas the cosmic X-ray background was ignored due to its negligible contribution. 
We note that \citet{Kammoun2025} used data taken during the time with cut-off rigidities (COR) greater than 6, but, following the canonical reduction, this work did not apply such a selection. 
The effect of the COR selection on the results is discussed in the Appendix of \citet{Kammoun2025}.
For Xtend, we extracted source and background spectra from 2\arcmin $\times$ 5\arcmin rectangular regions, excluding the source region for the background one. 
The net exposure times left for Resolve and Xtend are 111.8 and 83.2~ks, respectively. 
By using the \texttt{rslmkrmf} program, an X-large-sized redistribution matrix file (RMF) for Resolve was generated. 
Similarly, for Xtend, an RMF was produced based on the \texttt{xtdrmf} tool. 
Auxiliary response files (ARFs) for an on-axis point source were 
generated using the \texttt{xaarfgen} tool for both Resolve and Xtend. 

We analyzed NuSTAR data provided by the team based on the NuSTAR Data Analysis Software \texttt{NUSTARDAS} (version 2.1.4) and dedicated calibration files (version 20241104). 
The data were calibrated and cleaned by using the standard pipeline \texttt{nupipeline}. 
We then extracted source and background data from circular regions with radii of 75${^{\prime\prime}}$ and 90${^{\prime\prime}}$, respectively. 
Consequently, 
net exposure times of 53.4 and 52.9~ks are left for the two focal plane modules A and B (FPMA and FPMB), respectively.

\section{Spectral analysis}\label{sec:spec_ana}
For spectral analyses, we produced 2--10~keV Resolve, 3--10~keV Xtend, and 3--60~keV NuSTAR (FPMA and FPMB) source spectra averaged over the respective entire exposures. 
Taking the average is validated, by examining the light curves of the four detectors in the respective energy bands, no significant flux variability was found during our observations. For Xtend, we note that the energy range below 3~keV was ignored due to calibration uncertainties. 
While background events were subtracted from the Xtend and NuSTAR source spectra, we incorporated NXB contribution to the Resolve spectrum by simultaneously fitting an NXB model to the NXB and source spectra. 
By using the \texttt{grppha} tool, each X-ray spectrum was binned so that each bin includes at least one count. 
Any fitting was carried out via XSPEC version 12.15.0 \citep{Arnaud1996}, and best-fitting values were estimated based on the $C$-statistic \citep{Cash1979}. 
Throughout this work, uncertainties of fitted parameters are presented at the 90\% confidence level ($\Delta C$ = 2.706 for one parameter of interest), and we present all line energies in the rest frame.
When necessary, the metal abundances were assumed to be the solar values of \citet{Lodders2009}.

\subsection{Continuum Emission Determination} \label{sec:continuum}
Robust constraints on the continuum spectrum are a basis to identify emission lines. 
For this purpose, we focused on all four 
spectra, excluding the 5.8--7.5~keV band to ignore iron emission lines. 
The spectra were simultaneously fitted with a cutoff power-law model including Galactic absorption and a cross-calibration constant. 
We note that, although the neutral Fe K$\alpha$ emission is present, associated reflection continuum emission was ignored due to its weak contribution \citep{Kammoun2025}.
The Galactic absorbing hydrogen column density was fixed to $1.08\times10^{20}$~cm$^{-2}$ \citep{HI4PI2016}. 
The cross-calibration constant was fixed to unity for Resolve, 
while those for the other instruments were allowed to vary freely. 
Consequently, the best-fitting power-law photon index and high-energy cutoff were constrained to be $\Gamma = 1.74 \pm 0.02$ and $E_{\rm cut}=167^{+68}_{-38}$~keV with a $C$-stat./d.o.f. of 17047/17219. 
These values are consistent with those reported in \citet{Kammoun2025}.
The cross-calibration constants for the Xtend, FPMA, and FPMB spectra resulted in $1.01\pm0.01$, $1.05\pm0.01$, and $1.04\pm 0.01$, respectively. 
An estimated 2--10~keV flux is $F_{2-10\ \mathrm{keV}} = (4.87\pm0.02)\times10^{-11}$~erg~cm$^{-2}$~s$^{-1}$, corresponding to a 2--10~keV luminosity of $L_{2-10~\mathrm{keV}}= (3.67\pm0.01)\times10^{42}$~erg~s$^{-1}$.
As reported by \citet{Kammoun2025}, this flux level is $\sim2$ times higher than that observed with Chandra/HETG \citep{Bianchi2008}.
In subsequent spectral fittings, $\Gamma$ and $E_{\rm cut}$ of the best-fitting cutoff power-law model were adopted to reproduce the continuum emission, unless otherwise stated.

\subsection{Phenomenological models} \label{ssec:31} \label{sec:pheno_model}
\begin{figure*}[!h]
 \begin{center}
 \includegraphics[width=1\textwidth]{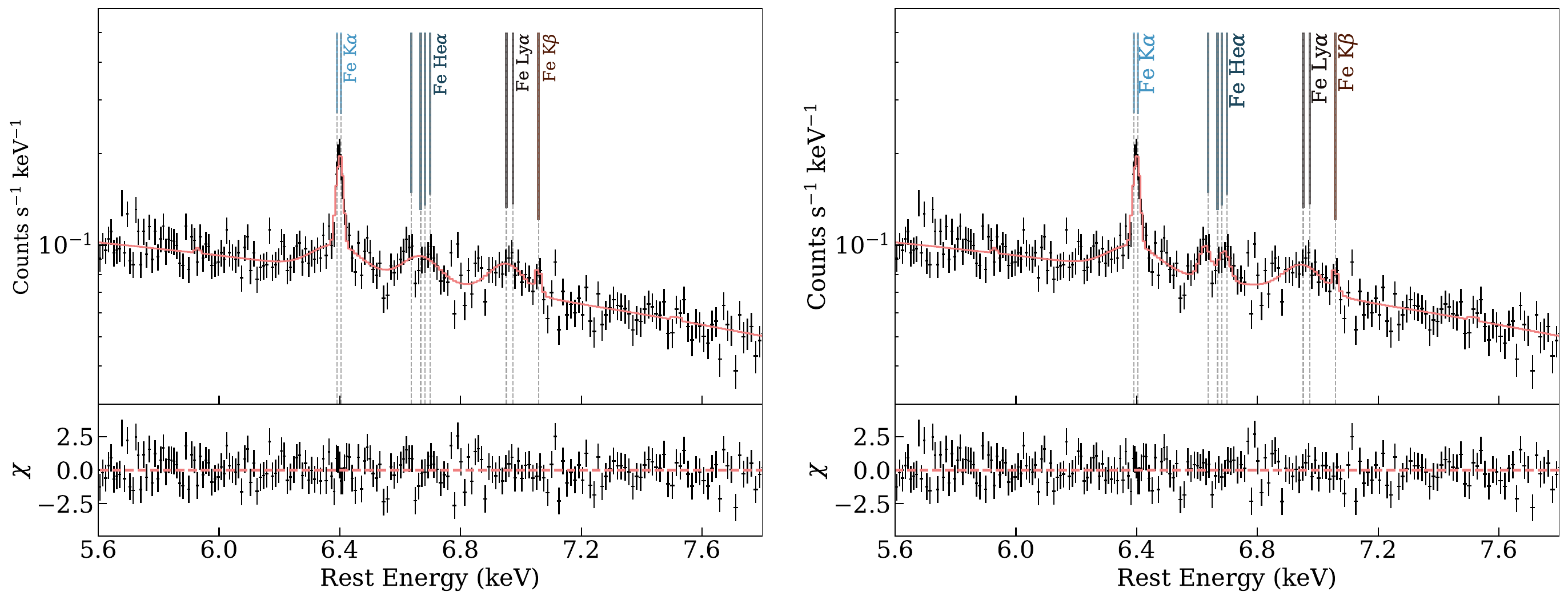}
 \end{center}
 \caption{
 Phenomenological fit results for the Resolve spectrum (\S\ref{sec:pheno_model}). 
 The left panel shows the fit result in which the velocity widths and redshifts of the Fe He$\alpha$ and Fe Ly$\alpha$ lines are tied together, while the right panel shows the one where these parameters are allowed to vary independently. Black crosses indicate the observed Resolve data, and the sum of all model components is shown as a pink line. The lower panel shows residuals of $\chi$ = (data $-$ model)$/$(1$\sigma$ error). The dashed vertical lines mark the rest-frame energies of Fe K$\alpha$, Fe K$\beta$, Fe He$\alpha$, and Fe Ly$\alpha$.
 {Alt text: Two panels showing phenomenological fits to the Resolve X-ray spectrum, with tied velocity widths and redshifts of the Fe He$\alpha$ and Fe Ly$\alpha$ lines in the left panel and untied these parameters in the right panel.}
 }
 \label{fig:rsl_pheno_fit}
\end{figure*}
Before applying physically motivated plasma models, we attempted to phenomenologically characterize Fe {\sc xxv} He$\alpha$ and Fe {\sc xxvi} Ly$\alpha$ emission in the Resolve spectrum with  Gaussian functions. 
Our phenomenological model assumed that the He$\alpha$ resonance line (w at 6.700~keV), intercombination lines (x at 6.682 and y at 6.668~keV), and forbidden line (z at 6.637~keV) share a line width and an energy shift, both of which were then treated as free parameters.
The normalizations of the four lines were allowed to vary independently. 
The Ly$\alpha_1$ and Ly$\alpha_2$ lines (6.973 and 6.952~keV, respectively) were modeled with the same line width and shift as the Fe He$\alpha$ lines. 
The normalization of the Ly$\alpha_1$ line was treated as a free parameter, whereas that of the Ly$\alpha_2$ line was fixed to half that of the Ly$\alpha_1$ line, as expected for optically thin plasma.
For the Fe K$\alpha$ emission, we considered both narrow and broad components, whereas the Fe K$\beta$ emission was included only for the narrow component, following the approach of \citet{Kammoun2025}. 
For simplicity, we modeled both the K$\alpha$ and K$\beta$ lines with Gaussian functions. 
The line width and energy shift of the Fe K$\alpha_1$ line (6.404~keV) were left as free parameters and tied to those of the Fe K$\alpha_2$ line (6.391~keV). 
The normalization of the K$\alpha_1$ line was left free, while that of the K$\alpha_2$ line was fixed to its one half. 
The Fe K$\beta$ (7.058~keV) line was modeled with the same line width and energy shift as the narrow K$\alpha$ lines, and its normalization was allowed to vary freely. 

The Resolve spectrum was fitted with the model consisting of the above-described Gaussian profiles for the iron lines plus a cutoff power-law component with $\Gamma \approx 1.74$ and $E_{\rm cut} \approx 167$~keV in \S\ref{sec:continuum}. 
The best-fitting result shown in the left panel of Figure~\ref{fig:rsl_pheno_fit} was obtained with $C$-stat./d.o.f. = 17024/17293. 
For the neutral iron emission lines, the narrow and broad components have velocity widths of $v_\sigma = 360^{+170}_{-120}$~km~s$^{-1}$ and $v_\sigma = 3570^{+2230}_{-1300}$~km~s$^{-1}$, respectively, consistent with the narrow and broad components reported by \citet{Kammoun2025}.
The highly ionized iron lines were fitted with a line width of $v_\sigma=2460^{+800}_{-1140}$~km~s$^{-1}$ and a redshift of $z=0.0020^{+0.0034}_{-0.0030}$ after correcting for the systemic redshift of $z = 0.005839$.
The normalization of the Fe Ly$\alpha_1$ line was constrained to be $0.97^{+0.31}_{-0.35}\times10^{-5}$~ph~s$^{-1}$~cm$^{-2}$. 
For the He$\alpha$ w, x, y, and z lines, the 90\% upper limits are estimated to be $2.27\times10^{-5}$, $1.42\times10^{-5}$, $1.45\times10^{-5}$ and $1.10\times10^{-5}$~ph~s$^{-1}$~cm$^{-2}$, respectively.
These upper limits do not imply that the data allow all four lines to be absent simultaneously, and are due to strong degeneracy between the lines.
\begin{table}[!b]
\caption{Parameters obtained from fitting the phenomenological models.}
\begin{center}
\begin{tabular}{lc}
\noindent Tied $v_{\sigma}$ and $z$ with Fe He$\alpha$ and Ly$\alpha$ & \\
\hline
\hline
Parameter & Value\\
\hline
Cutoff PL norm\footnotemark[$*$] & $1.27^{+0.01}_{-0.01}$ \\
\\
Fe He$\alpha$ $v_{\sigma}$ (km~s$^{-1}$) & $2460^{+800}_{-1140}$ \\
Fe He$\alpha$ $z$\footnotemark[$\dagger$] & $0.0020^{+0.0034}_{-0.0030}$ \\
Fe He$\alpha$ (w) norm (10$^{-5}$ ph~s$^{-1}$~cm$^{-2}$) & $<$ 2.27 \\
Fe He$\alpha$ (x) norm & $<$ 1.42 \\
Fe He$\alpha$ (y) norm & $<$ 1.45 \\
Fe He$\alpha$ (z) norm & $<$ 1.10 \\
\\
Fe Ly$\alpha$ $v_{\sigma}$ (km~s$^{-1}$) & $=$ Fe He$\alpha$ \\
Fe Ly$\alpha$ $z$\footnotemark[$\dagger$] & $=$ Fe He$\alpha$ \\
Fe Ly$\alpha_1$ norm (10$^{-5}$ ph~s$^{-1}$~cm$^{-2}$) & $0.97^{+0.31}_{-0.35}$ \\
Fe Ly$\alpha_2$ norm & $= 0.5\times \mathrm{Ly\alpha_1}$ \\
\hline
$C$-stat./d.o.f. & 17024/17293 \\
\hline 
\\
\noindent Independent $v_{\sigma}$ and $z$ for Fe He$\alpha$ and Ly$\alpha$ & \\
\hline
\hline
Parameter & Value\\
\hline
Cutoff PL norm\footnotemark[$*$] & $1.27^{+0.01}_{-0.01}$ \\
\\
Fe He$\alpha$ $v_{\sigma}$ (km~s$^{-1}$) & $790^{+370}_{-240}$ \\
Fe He$\alpha$ $z$\footnotemark[$\dagger$] & $0.0015^{+0.0012}_{-0.0013}$ \\
Fe He$\alpha$ (w) norm (10$^{-5}$ ph~s$^{-1}$~cm$^{-2}$) & $0.54^{+0.65}_{-0.25}$ \\
Fe He$\alpha$ (x) norm & $<$ 0.20 \\
Fe He$\alpha$ (y) norm & $<$ 0.15 \\
Fe He$\alpha$ (z) norm & $0.60^{+0.24}_{-0.23}$ \\
\\
Fe Ly$\alpha$ $v_{\sigma}$ (km~s$^{-1}$) & $2610^{+1700}_{-1580}$ \\
Fe Ly$\alpha$ $z$\footnotemark[$\dagger$] & $0.0024^{+0.0033}_{-0.0034}$ \\
Fe Ly$\alpha_1$ norm (10$^{-5}$ ph~s$^{-1}$~cm$^{-2}$) & $1.02^{+0.43}_{-0.35}$ \\
Fe Ly$\alpha_2$ norm & $= 0.5\times \mathrm{Ly\alpha_1}$ \\
\hline
$C$-stat./d.o.f. & 17018/17291 \\
\hline
\end{tabular}
\label{tab:phenomenologicalfittable}
\end{center}
\begin{tabnote}
\footnotemark[$*$] The normalization of the cutoff power-law model in units of $10^{-2}$~photons~keV$^{-1}$~cm$^{-2}$~s$^{-1}$ at 1~keV. \\
\footnotemark[$\dagger$] The redshift parameter $z$ with respective to the source rest frame ($z$ = 0.005839). \\
\end{tabnote}
\end{table}

Although the above fit tied the line width and redshift of the He$\alpha$ and Ly$\alpha$ complex lines, we also tested a model in which their widths and redshifts can vary independently.
The right panel of Figure~\ref{fig:rsl_pheno_fit} shows the result. 
The fit was slightly improved, yielding a $C$-stat./d.o.f. of 17018/17291 ($\Delta C = 6$). 
While this improvement corresponds to only $\sim 2\sigma$ and cannot be considered statistically significant,
we found that the Fe He$\alpha$ and Ly$\alpha$ lines may have different widths with a consistent redshift; $v_\sigma = 790^{+370}_{-240}$~km~s$^{-1}$ and $z = 0.0015^{+0.0012}_{-0.0013}$ for the Fe~He$\alpha$ lines, and $v_\sigma = 2610^{+1700}_{-1580}$~km~s$^{-1}$ and $z = 0.0024^{+0.0033}_{-0.0034}$ for the Fe Ly$\alpha$ lines. 
A similar argument was made also in \cite{Kammoun2025}.
The result indicates that the He-like and H-like iron emissions may trace different gas structures.
In this case, the normalizations of the He$\alpha$ w, z and Fe Ly$\alpha_1$ lines were constrained to be $0.54^{+0.65}_{-0.25}\times10^{-5}$, $0.60^{+0.24}_{-0.23}\times10^{-5}$ and $1.02^{+0.43}_{-0.35}\times10^{-5}$~ph~s$^{-1}$~cm$^{-2}$, respectively. 
The He$\alpha$ x and y lines were, however, not significantly detected, however;  
the upper limits of the x and y lines were estimated to be $0.20\times10^{-5}$ and $0.15\times10^{-5}$~ph~s$^{-1}$~cm$^{-2}$, respectively.

\subsection{Photoionization models} \label{ssec:32} \label{sec:photo_model}
\subsubsection{Creation of a photoionization model} \label{ssec:311} \label{sec:photo_create} 
We investigate whether the Fe He$\alpha$ and Ly$\alpha$ emission lines originate from photoionized plasma using  the \texttt{pion} model \citep{Mehdipour2016}. 
We utilized \texttt{pion} in the X-ray spectral analysis software SPEX v3.08.01 \citep{Kaastra1996}. 
The model solves for the ionization equilibrium of gas exposed to given ionizing radiation, computes the ionization fraction of each ion, and derives line and continuum emission from a photoionized gas slab. 
The \texttt{pion} model has the advantage that, during the fit, it self-consistently recalculates the line and continuum emission whenever the incident spectrum is changed, but at the cost of increased computational expense.
In the present study, because our goal is to survey a wide range of parameters describing the photoionized gas, we therefore fixed the incident continuum.
For this fixed illuminating spectrum, we precomputed a grid of photoionized spectra over the relevant parameter space and compiled them into a table model, which allows us to perform the fits much more efficiently.
\begin{figure}[!t]
 \begin{center}
  \includegraphics[width=1.0\columnwidth]{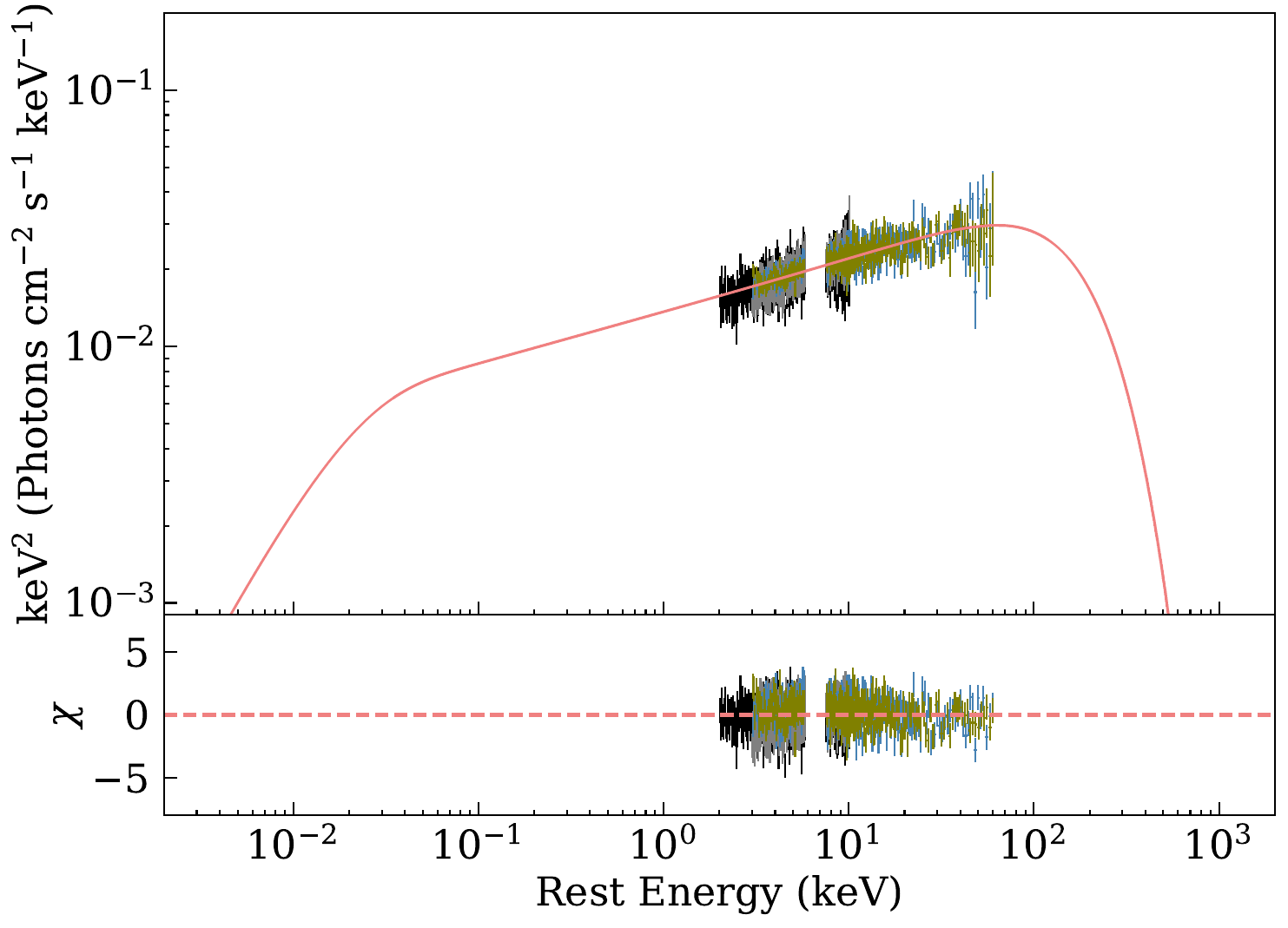}
 \end{center}
 \caption{Input SED model of \texttt{nthcomp} (pink) used for generating the \texttt{pion} table model.
 In the lower panel, residuals, $\chi$ = (data -- model)$/$(1$\sigma$ error), are shown. 
 Black, gray, blue, and green crosses represent the Resolve, Xtend, FPMA, and FPMB data, respectively.
 {Alt text: A panel showing the input spectral energy distribution model of nthcomp used to generate the pion table model. }
 }
 \label{fig:input_sed}
\end{figure}

As the incident spectral energy distribution (SED) for the photoionization calculations, we adopted a thermal Comptonization continuum modeled with \texttt{nthcomp}. 
The seed-photon temperature $kT_{\rm bb}$ is difficult to constrain observationally, and the electron temperature $kT_{\rm e}$ can be degenerate with the photon index. 
We thus assumed $kT_{\rm bb} = 10$~eV and $kT_{\rm e} = 100$~keV. 
The former value is motivated by the finding of \cite{Kammoun2025} that the inner radius of the accretion disk extends at least to $\sim 100\,r_{\rm g}$. The disk blackbody temperature may thus be $\gtrsim$ 1~eV. 
The latter temperature is based on the high-energy cutoff of $\approx 170$~keV constrained from the joint XRISM and NuSTAR spectra in subsection~\ref{sec:continuum}. 
It roughly corresponds to an electron temperature of several tens of keV.
Even when adopting an electron temperature of $kT_{\rm e} = 50$~keV, the fit results are largely unaffected.
The remaining parameter, the photon index, was then determined by fitting the \texttt{nthcomp} model to the XRISM and NuSTAR spectra, excluding the iron band (5.8--7.5~keV), with only the normalization left free.
As shown in Figure~\ref{fig:input_sed}, this fit yields $\Gamma \sim 1.8$, and we adopted $\Gamma = 1.8$ in the subsequent photoionization modeling. 

We note that NGC\,7213 may exhibit a weak big blue bump \citep{Starling2005,Lobban2010}, and we thus examined how largely including a disk component in the incident SED model affects our discussion. 
To estimate the disk contribution,
we additionally used optical and UV data taken by quasi-simultaneous XMM-Newton Optical Monitor (OM) observations during the XRISM observing period.
V, B, U, UVW1, UVM2, and UVW2 data were taken, and 
after reprocessing the data, we measured 
fluxes using the \texttt{omphotom} task. 
In the measurements, source and background regions were set to 
a circular region with a radius of 5.8\arcsec\,
and an annulus between 7.5\arcsec and 12.2\arcsec, respectively. 
The regions correspond to those typically used for measuring 
the flux of a point source, and 
by adopting them, we attempted to subtract host galaxy emission as well. 
As host galaxy emission may remain however, 
we conservatively used them as upper limits for the AGN emission. 
Then, adopting \texttt{tbabs*redden(diskbb*nthcomp)}, 
we determined an SED spectrum, which includes the maximum disk contribution allowed by the data. In the determination, 
while the photo-electronic absorption model (\texttt{tbabs}) adopted the galactic column density of $1.08\times10^{20}$ cm$^{-2}$, 
the color E(B-V) in the extinction model \texttt{redden} was set to $\approx$ 0.02 equivalent to the column density.
As a result, the temperature of the innermost disk radius was estimated to be $\approx$ 0.5 eV. 
With the model, we generated the \texttt{pion} table model 
and confirmed that the derived ionization and kinematic parameters are consistent with those obtained with the baseline \texttt{nthcomp}-only SED within the uncertainties. 

With the incident spectrum fixed (Figure~\ref{fig:input_sed}), we computed emission spectra over a wide grid of photoionized gas parameters and compiled the results into a table model for use in the spectral fitting.
The parameters include the ionization parameter ($\xi$), the hydrogen density ($n_{\rm H}$), the hydrogen column density ($N_{\rm H}$), and the root mean square velocity width ($v_\sigma$). 
The ionization parameter is here defined as $\xi = L_{\rm ion}/(n_{\rm H} R^2)$, where $L_{\rm ion}$ is the ionizing luminosity at energies above 13.6~eV and $R$ is the distance from the ionizing source. 
To reduce the number of parameters, we fixed the density at $n_{\rm H} = 1\times10^8\,\mathrm{cm^{-3}}$. 
Although the value is discussed and validated later in subsection~\ref{ssec:312}, 
we note that the relative intensity ratios of Fe He$\alpha$ and Ly$\alpha$ emission lines are not so sensitive to the density around the value. 
This is natural as the critical density for Fe {\sc xxv} is $\sim 10^{16}$~cm$^{-3}$ \citep{Blumenthal1972}.
Then, by varying the ionization parameter ($2\leq \log(\xi/\mathrm{erg\,cm\,s^{-1}})\leq8$ in steps of 0.2), the column density ($19 \leq \log (N_{\rm\,H}/\mathrm{cm^{-2}})\leq25$ in steps of 0.2), and the velocity width ($-4 \leq \log(v_{\sigma}/c)\leq-1$ in steps of 0.2), we generated 15,376 emission spectra.  
Each spectrum was sampled into 65,536 logarithmically spaced energy bins between 1~eV and 1~MeV. 
This binning gives an energy width of $\sim 1.5$~eV around 7~keV, fine enough to model the high-resolution Resolve spectrum. 
Furthermore, each was rescaled so that its normalization \texttt{norm$_{\rm pion}$} parameter represents $L_{\rm bol}/(4\pi D^2)\Omega$,
where $D$ is the distance from the photoionized gas to the observer ($\sim $ distance from the continuum source to the observer) and $0\leq\Omega \leq 1$ is the ratio of solid angle to $4\pi$.
Finally, by compiling the generated spectra, we obtained a \texttt{pion} table model.
This was used as an additive model in XSPEC.

\subsubsection{Spectral fitting with a one-zone photoionization model} \label{ssec:312}
\begin{figure*}[!ht]
 \begin{center}
 \includegraphics[width=.95\textwidth]{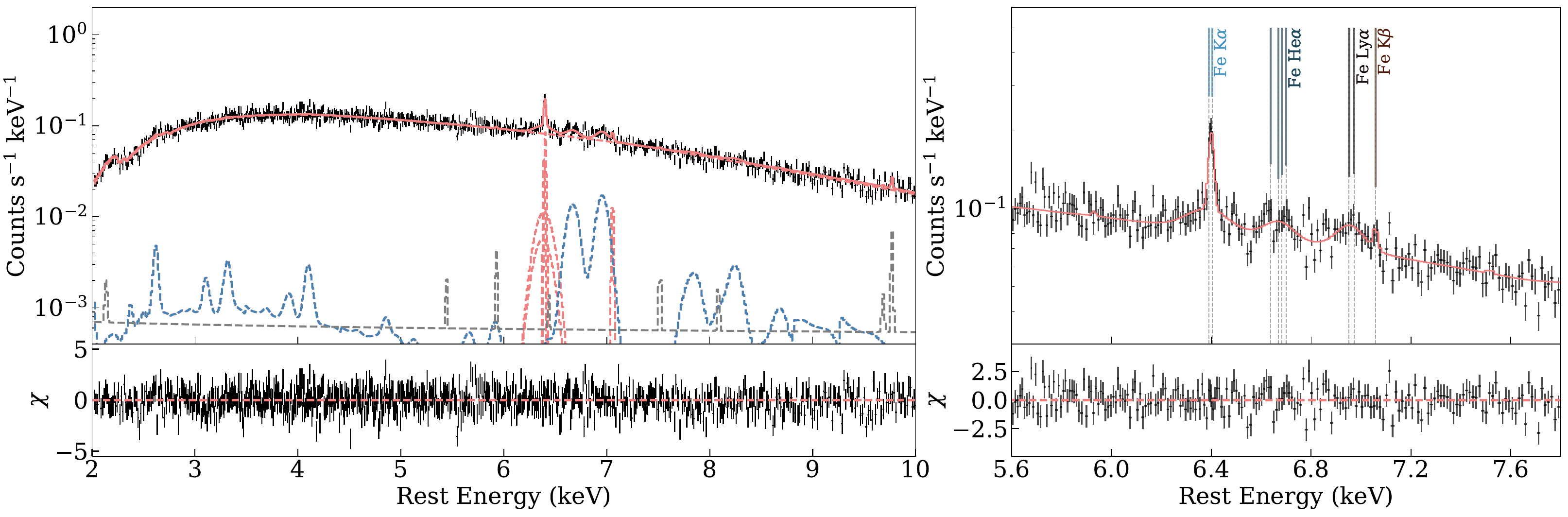}
 \includegraphics[width=.95\textwidth]{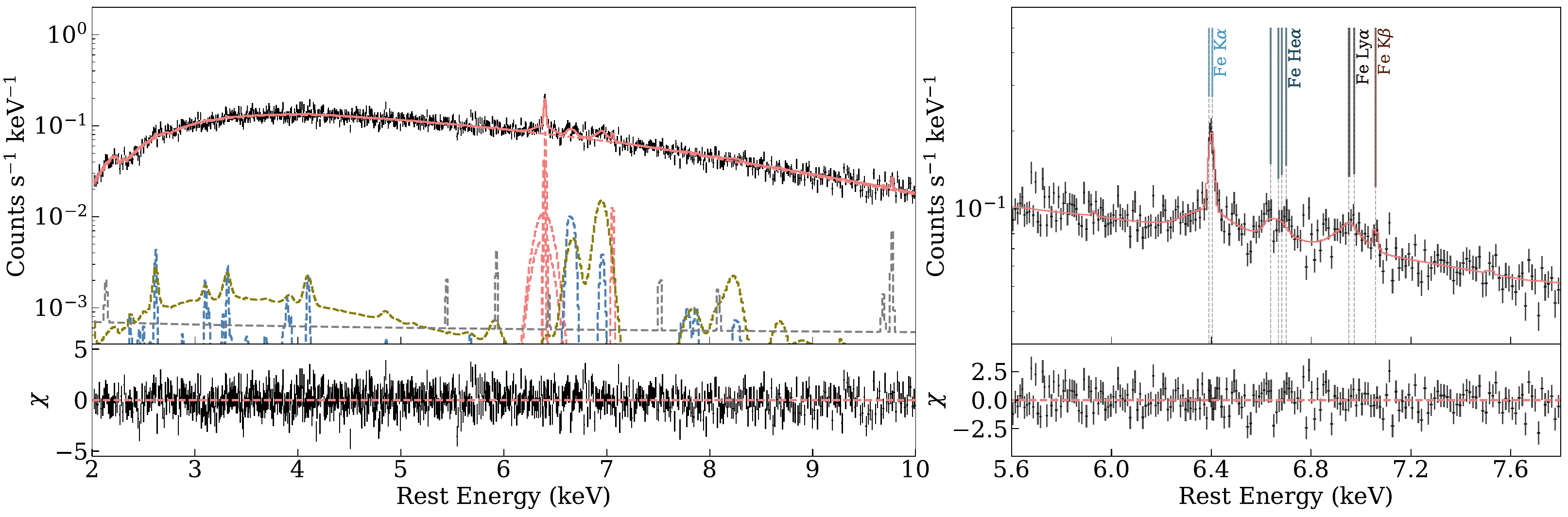}   
 \end{center}
 \caption{
 Top left: Resolve spectrum (black crosses) and the best-fitting model (pink solid line) adopting the single \texttt{pion} component. The pink dashed lines represent the cutoff power-law component, narrow and broad Fe K$\alpha$ lines, and Fe K$\beta$ line. The blue dashed line corresponds to the \texttt{pion} component. The gray dashed line indicates the NXB. 
  The lower panel shows residuals of $\chi$ = (data $-$ model)$/$(1$\sigma$ error). 
 Top right: Enlarged view of the iron emission lines, where the dashed vertical lines mark the rest-frame energies of Fe K$\alpha$, Fe K$\beta$, Fe He$\alpha$, and Fe Ly$\alpha$.
 Bottom left and right: Same as the top panels, except for the model adopting two \texttt{pion} components. The additional green dashed line indicates the second \texttt{pion} component. 
 {Alt text: Figures showing spectral fits using one and two component photoionization models.}
 }\label{fig:rsl_pion_repro_fit}
\end{figure*}

\begin{figure}[!h]
 \begin{center}
 \includegraphics[width=1.0\columnwidth]{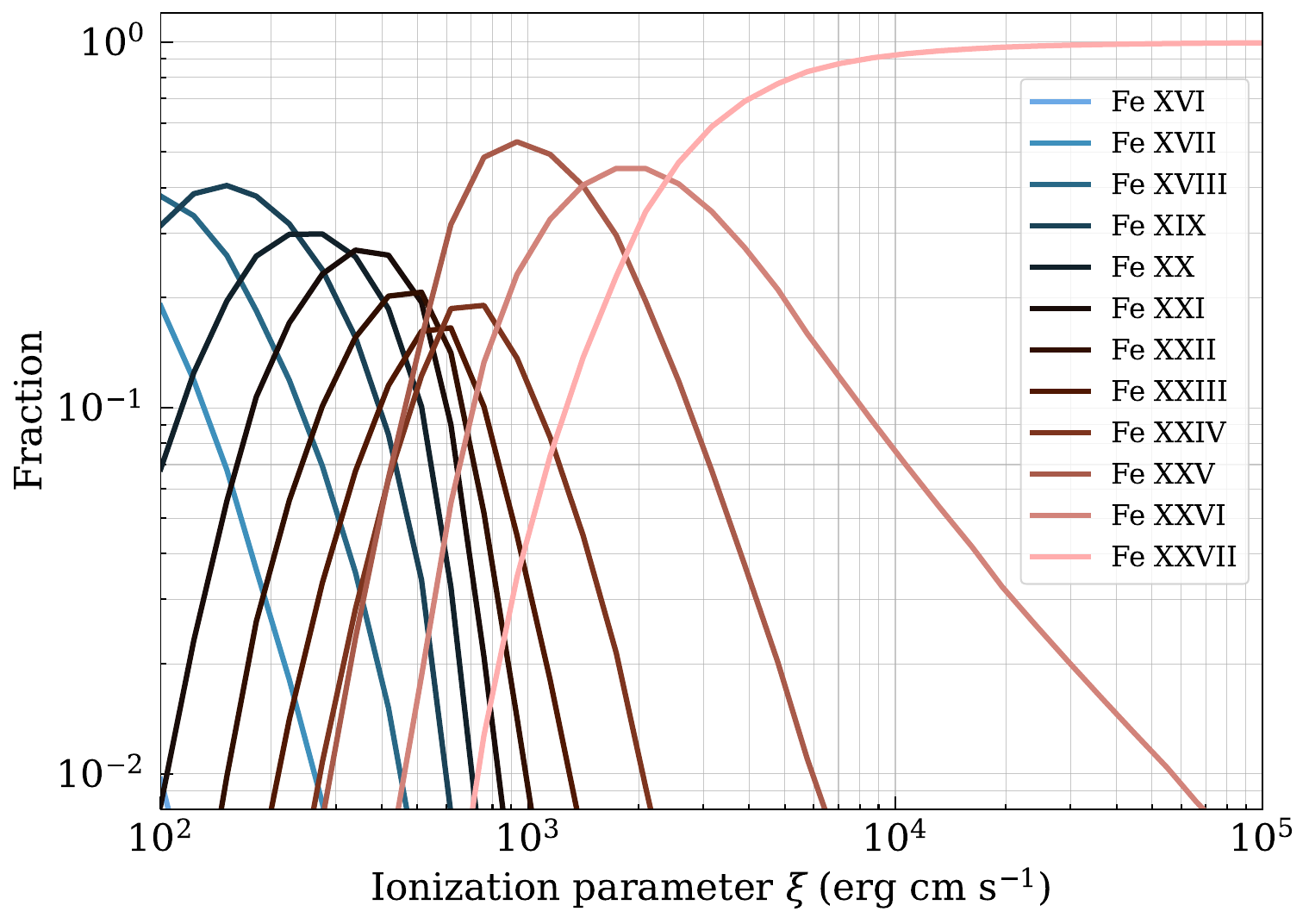}
 \end{center}
 \caption{
 Ion fractions of iron ions from Fe {\sc xvi} and higher calculated with our photoionization model. The horizontal axis shows the ionization parameter, and the vertical axis shows the ionic fraction.
 {Alt text: A plot showing ion fractions of iron ions from Fe XVI and higher calculated with a photoionization model, as a function of the ionization parameter.}
 }
 \label{fig:pion_ion_frac}
\end{figure}

We fitted the Resolve spectrum using the created \texttt{pion} model (\S\ref{sec:photo_create}) together with a cutoff power-law component and neutral iron emission lines. This corresponds to a one-zone photoionization model. 
Among the \texttt{pion} parameters, the ionization parameter, normalization, velocity width, and redshift were left free, but the column density ($N_\mathrm{H}$) was fixed at $1\times10^{24}$~cm$^{-2}$, as it cannot be constrained well. 
As later discussed, even if we change it, results are not affected significantly. 
The photon index and the cutoff energy of the power-law component 
were fixed to those obtained from the XRISM and NuSTAR joint fit (\S\ref{sec:continuum}).   
The neutral lines were modeled in the same way as in \S\ref{sec:pheno_model}, except that the K$\beta$-to-K$\alpha$ flux ratio was fixed to 0.13. 
Figure~\ref{fig:rsl_pion_repro_fit} shows the best-fitting result obtained with $\log \xi = 3.36^{+0.15}_{-0.10}$, at which He-like and H-like ions are comparably dominant, followed by Li-like ions (Figure~\ref{fig:pion_ion_frac}).
It is seen that the \texttt{pion} component reproduces the He$\alpha$ lines like the Gaussian model assuming a single velocity for the He$\alpha$ and Ly$\alpha$ lines. 
Accordingly, the constrained velocity width ($2450^{+860}_{-730}$~km~s$^{-1}$) and redshift ($0.0014^{+0.0027}_{-0.0028}$) are  consistent with those obtained from the Gaussian modeling ($2460^{+800}_{-1140}$~km~s$^{-1}$ and $0.0020^{+0.0034}_{-0.0030}$).
All fitted values are summarized in Table~\ref{tab:pionfittable}. 
\begin{table}[!b]
\caption{Parameters obtained from fitting the photoionization models.}
\begin{center}
\begin{tabular}{lc}
\noindent One-zone \texttt{pion} model & \\
\hline
\hline
Parameter & Value\\
\hline
Cutoff PL norm\footnotemark[$*$] & $1.26^{+0.01}_{-0.01}$ \\
\\
Emis. $\log \xi$ (erg~cm~s$^{-1}$) & $3.36^{+0.15}_{-0.10}$ \\
Emis. $N_\mathrm{H}$ ($10^{24}$~cm$^{-2}$) & 1 (fix) \\
Emis. $v_{\sigma}$ (km~s$^{-1}$) & $2450^{+850}_{-730}$ \\
Emis. $z$\footnotemark[$\dagger$] & $0.0014^{+0.0027}_{-0.0028}$ \\
Emis. norm ($10^{-11}$~erg~cm$^{-2}$~s$^{-1}$) & $3.23^{+1.67}_{-0.97}$ \\
\hline
$C$-stat./d.o.f. & 17029/17297 \\
\hline 
\\
\noindent Two-zone \texttt{pion} model & \\
\hline
\hline
Parameter & Value\\
\hline
Cutoff PL norm\footnotemark[$*$] & $1.26^{+0.01}_{-0.04}$ \\
\\
Emis. 1 $\log \xi$ (erg~cm~s$^{-1}$) & $3.00^{\ddagger}_{-0.25}$ \\
Emis. 1 $N_\mathrm{H}$ ($10^{24}$~cm$^{-2}$) & 1 (fix) \\
Emis. 1 $v_{\sigma}$ (km~s$^{-1}$) & 790 (fix) \\
Emis. 1 $z$\footnotemark[$\dagger$] & 0.0015 (fix) \\
Emis. 1 norm ($10^{-11}$~erg~cm$^{-2}$~s$^{-1}$) & $0.47^{+0.54}_{-0.46}$ \\
\\
Emis. 2 $\log \xi$ (erg~cm~s$^{-1}$) & $3.58^{\ddagger}_{-0.24}$ \\
Emis. 2 $N_\mathrm{H}$ ($10^{24}$~cm$^{-2}$) & 1 (fix) \\
Emis. 2 $v_{\sigma}$ (km~s$^{-1}$) & 2610 (fix) \\
Emis. 2 $z$\footnotemark[$\dagger$] & 0.0024 (fix) \\
Emis. 2 norm ($10^{-11}$\,erg\,cm$^{-2}$~s$^{-1}$) & $4.26^{+11.85}_{-2.19}$ \\
\hline
$C$-stat./d.o.f. & 17027/17297 \\
\hline
\end{tabular}
\label{tab:pionfittable}
\end{center}
\begin{tabnote}
\footnotemark[$*$] The normalization of the cutoff power-law model in units of $10^{-2}$~photons~keV$^{-1}$~cm$^{-2}$~s$^{-1}$ at 1~keV. \\
\footnotemark[$\dagger$] The redshift parameter $z$ with respective to the source rest frame ($z$ = 0.005839). \\
\footnotemark[$\ddagger$] Limit was not constrained. 
\end{tabnote}
\end{table}

The above fit confirms that the gas density of $n_{\rm H} = 1\times10^{8}\,\mathrm{cm^{-3}}$ assumed in the \texttt{pion} table model (\S\ref{ssec:311}) would be reasonable. 
A density can be estimated via the ionization parameter  
as $n_{\rm H} = L_{\mathrm{ion}} / \xi R^2$. 
While adopting the best-fitting value of $\log \xi=3.36$, we 
approximate $L_{\rm ion}$, or the AGN luminosity at energies above 13.6~eV, and $R$. 
We use $L_{\mathrm{bol}}$ as a proxy for $L_{\rm ion}$. 
The luminosity can be estimated from the observed 2--10~keV luminosity by adopting a bolometric correction factor of 10 \citep{Duras2020}, yielding $L_{\mathrm{bol}} \sim 4\times10^{43}$~erg~s$^{-1}$. 
For the radius $R$, although the observed velocity width may originate from various physical processes (e.g., outflows), we assume that the emission arises from gas located in the disk plane or a structure associated with the disk, and that the observed line broadening is due to Keplerian rotation given by
$r_{\mathrm{kep}} = GM/v_{\mathrm{rot}}^{2}$ (the radius expected for the gravitational constant of $G$, the central mass of $M$, and the rotational velocity $v_{\rm rot}$).
We then use $r_{\mathrm{kep}}$ as the radius $R$; thus, $R \sim GM/v^2_{\rm rot}$. 
To connect the Keplerian rotational velocity ($v_{\rm rot}$) and the observed line-of-sight velocity ($v_{\rm obs}$), we adopt $v_{\mathrm{rot}}=v_{\mathrm{obs}}/\sin\,i$, where $i$ is the inclination angle measured from the axis perpendicular to the disk plane. 
By collecting the necessary information: a possible inclination angle of $i=11^{\circ}$, estimated by modeling the broad Fe K$\alpha$ emission line with a relativistic accretion disk model \citep{Kammoun2025},
the velocity width of $v_{\mathrm{obs}}=2450$~km~s$^{-1}$, and $M_{\rm BH} = 10^{8}~M_{\odot}$ \citep{Woo2002}, $R$ is estimated to be $\sim 3\times10^{-3}$ pc, or $\sim 5\times10^{2}~r_{\mathrm{g}}$, where $r_{\mathrm{g}} = GM/c^{2}$ for the speed of light $c$. 
Consequently, the hydrogen density can be estimated to be $n_{\rm H}\sim 2\times10^{8}$~cm$^{-3}$, confirming that our assumption is reasonable.

Based on the almost same way adopted for validating the number density, we also confirm that the fixed hydrogen column density of $10^{24}$~cm$^{-2}$ seems to be reasonable. 
A hydrogen column density is estimated as $N_{\rm H} \sim n_{\rm H} \Delta R$, where $\Delta R$ is the length along the line of sight.
Here, we roughly assume $\Delta R \sim r_{\rm kep}$; thus, we eventually calculate 
\begin{equation} \label{eqn:nh}
    N_{\mathrm{H}} \sim \frac{L_{\mathrm{ion}}}{\xi \, r_{\mathrm{kep}}}.
\end{equation}
Using the previously used values ($v_{\rm obs} = 2450$~km~s$^{-1}$, $i = 11^\circ$, $L_{\rm ion} \sim 4\times10^{43}$~erg~s$^{-1}$, and 
$M = 10^{8}$\,$M_\odot$), a column density is estimated to be 
$\sim 2\times10^{24}$~cm$^{-2}$, confirming that our assumption is reasonable.
We note that even if we adopt a different column density and refit the Resolve spectrum, the ionization parameter does not change much. This suggests that, for a self-consistent fit, $N_{\rm H} \sim 10^{24}$~cm$^{-2}$ is desired.

\subsubsection{Spectral fitting with a two-zone photoionization model} \label{sec:twozone_photo}

The tentative possibility that the Fe He$\alpha$
and Ly$\alpha$ lines have different velocity widths was presented in \S\ref{sec:pheno_model}, and, motivated by this, to make our investigation of the iron lines comprehensive, we added another pion component to the previous model, resulting in a two-zone model.
Then, we made each component represent one of the two velocity properties as well as the corresponding redshift. 
In its fit, the ionization parameter and normalization of the added \texttt{pion} component were allowed to vary independently of those of the other.  
The column densities of both components were fixed at $1 \times 10^{24}$~cm$^{-2}$, as it cannot be constrained. 
The bottom panel of Figure~\ref{fig:rsl_pion_repro_fit} shows the result, and the best-fitting parameters are listed in Table \ref{tab:pionfittable}. 
The figure shows that while the Ly$\alpha$ lines are predominantly explained by $\log \xi = 3.58$, the Fe He$\alpha$ lines have a larger contribution from the other less-photoionized plasma with $\log \xi = 3.00$. 
If the observed lines are broadened following the Keplerian rotation, the two-zone \texttt{pion} fit might suggest that the Fe Ly$\alpha$ emission is predominantly produced in an inner, more highly ionized region, while the He$\alpha$ emission originates from an outer, less ionized region. 
However, the added component, motivated by the Gaussian fitting, does not lead to a significant improvement in the fit and is not strongly required.

We note that, as proved for the one-zone model, the fixed number and column density ($n_{\rm H}$ and $N_{\rm H}$) are likely to be reasonable also for the two-zone model. The actual estimated values of $n_{\rm H}$ for the lowly and highly ionized components
are $10^{7} \sim 10^{8}$~cm$^{-3}$, while the column densities are $\sim5\times10^{23}$ and $\sim1\times10^{24}$~cm$^{-2}$, respectively.

\subsection{One- and two-zone collisional ionization models} \label{ssec:33}\label{sec:col_model}
\begin{figure*}[!t]
 \begin{center}
 \includegraphics[width=1\textwidth]{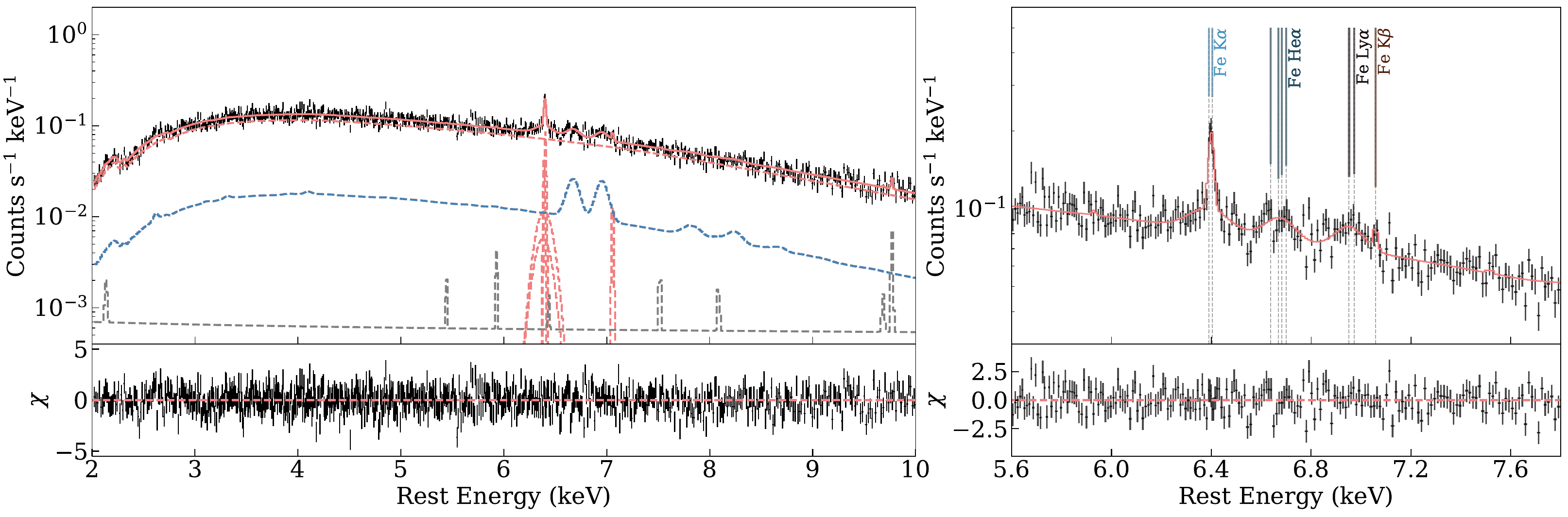}
 \includegraphics[width=1.0\textwidth]{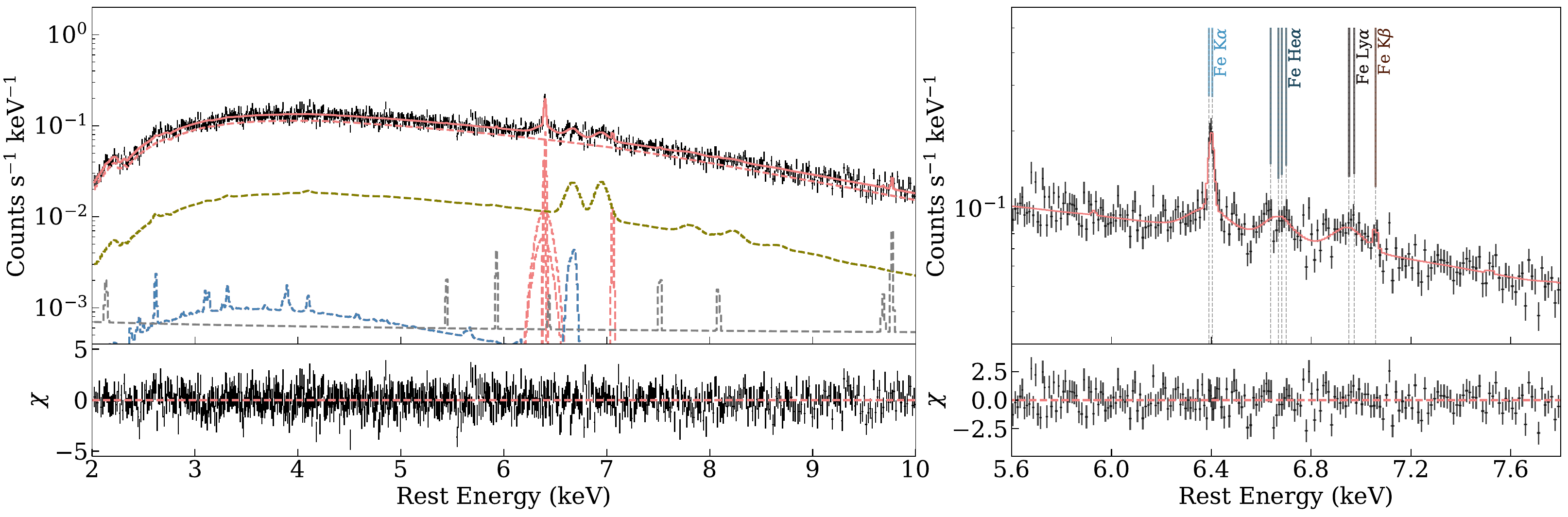}
 \end{center}
 \caption{
     Top left: The best-fitting model obtained with the one component \texttt{bapec} model. The pink dashed lines represent the cutoff power-law continuum together with the narrow and broad Fe K$\alpha$ lines and the Fe K$\beta$ line. 
     The blue and gray dashed lines correspond to the \texttt{bapec} and NXB components, respectively. 
     The sum of all these components is shown as the pink solid line. 
     Top right: Enlarged view of the iron emission lines, where the dashed vertical lines mark the rest-frame energies of Fe K$\alpha$, Fe K$\beta$, Fe He$\alpha$, and Fe Ly$\alpha$.
     Bottom left and right: Same as the top panels, except for the fitted model that considers two \texttt{bapec} components. 
     The additional \texttt{bapec}  component is plotted as the green dashed line. 
     {Alt text: Figures showing spectral fits using one and two component collisional ionization models.}
 }
 \label{fig:rsl_bapec_d_fit}
\end{figure*}

We also investigate the origin of the highly ionized iron lines using collisionally ionized plasma.
For this purpose, we considered the \texttt{bapec} model in XSPEC. 
It calculates emission from a collisionally ionized, optically thin thermal plasma, and applies a velocity broadening based on a Gaussian function.  
The model is parameterized with temperature, redshift, velocity width, and normalization. The normalization is defined as $\frac{10^{-14}}{4\pi[D_A(1+z)]^2}\int{n_{\rm e} n_{\rm H}dV}$, where $D_A$ is the angular diameter distance to the source, $dV$ is the volume element, and $n_{\rm e}$ and $n_{\rm H}$ are the electron and hydrogen number densities, respectively. 
$\int n_{\rm e} n_{\rm H} dV$ corresponds to the emission measure (EM). 
We combine the \texttt{bapec} model with the four parameters allowed to vary freely with neutral iron line and cutoff power-law components in the same way as in the photoionization model fits. 
With the total model, we then reproduced the Resolve spectrum with $C$-stat./d.o.f. = 17024/17297 (Figure~\ref{fig:rsl_bapec_d_fit}). 
The fitted temperature is $10.8^{+1.7}_{-1.5}$~keV, consistent with past estimates of $kT\sim10$~keV based on Suzaku and XMM-Newton \citep[e.g.,][]{Lobban2010,Emmanoulopoulos2013}. 
To reproduce the Fe He$\alpha$ and Ly$\alpha$ lines with comparable fluxes, such a high temperature appears to be required. 
The obtained velocity width ($2410^{+790}_{-680}$~km~s$^{-1}$) and redshift ($0.0016^{+0.0026}_{-0.0021}$) are consistent with those derived from the velocity-tied phenomenological model fit.
All best-fitting parameters are listed in Table~\ref{tab:bapecfittable}. 
\begin{table}[!b]
\caption{Parameters obtained from fitting the collisional ionization models.}
\begin{center}
\begin{tabular}{lc}
One-zone \texttt{bapec} model \\ 
\hline
\hline
 Parameter & Value\\
\hline
Cutoff PL norm\footnotemark[$*$] & $1.10^{+0.05}_{-0.05}$ \\
\\
Emis.  $kT$ (keV) & $10.8^{+1.7}_{-1.5}$ \\
Emis.  metal abundances & 1 (fix) \\
Emis.  $v_{\sigma}$ (km~s$^{-1}$) & $2410^{+790}_{-680}$ \\
Emis.  $z$ \footnotemark[$\dagger$] & $0.0016^{+0.0026}_{-0.0021}$ \\
Emis.  EM ($10^{64}$ cm$^{-3}$) & $2.92^{+0.79}_{-0.76}$ \\
\hline
$C$-stat./d.o.f. & 17024/17297 \\
\hline 
\\
Two-zone \texttt{bapec} model \\ 
\hline 
\hline 
 Parameter & Value\\
\hline
Cutoff PL norm\footnotemark[$*$] & $1.09^{+0.05}_{-0.06}$ \\
\\
Emis. 1 $kT$ (keV) & 3.0 (fix) \\
Emis. 1 metal abundances & 1 (fix) \\
Emis. 1 $v_{\sigma}$ (km~s$^{-1}$) & 790 (fix) \\
Emis. 1 $z$\footnotemark[$\dagger$] & 0.0015 (fix) \\
Emis. 1 EM ($10^{64}$ cm$^{-3}$) & $0.27^{+0.68}_{\ddagger}$ \\
\\
Emis. 2 $kT$ (keV) & $11.5^{+5.0}_{-2.0}$ \\
Emis. 2 metal abundances & 1 (fix) \\
Emis. 2 $v_{\sigma}$ (km~s$^{-1}$) & 2610 (fix) \\
Emis. 2 $z$\footnotemark[$\dagger$] & 0.0024 (fix) \\
Emis. 2 EM ($10^{64}$ cm$^{-3}$) & $2.98^{+0.83}_{-0.74}$ \\
\hline
$C$-stat./d.o.f. & 17025/17298 \\
\hline
\end{tabular}
\end{center}
\label{tab:bapecfittable}
\begin{tabnote}
\footnotemark[$*$] The normalization of the cutoff power-law model in units of $10^{-2}$~photons~keV$^{-1}$~cm$^{-2}$~s$^{-1}$ at 1~keV. \\
\footnotemark[$\dagger$] The redshift parameter $z$ with respective to the source rest frame ($z$ = 0.005839).\\
\footnotemark[$\ddagger$] Limit was not constrained.
\end{tabnote}
\end{table}

Considering that the Fe He$\alpha$ and Ly$\alpha$ lines may represent different velocity components (\S\ref{sec:pheno_model}), we added another \texttt{bapec} model to the previous one-zone collisional model and refitted the two-zone model to the Resolve spectrum. 
In the fit, each \texttt{bapec} component represents one of the two velocity components suggested by the Gaussian fit that allowed the He$\alpha$ and Ly$\alpha$ lines to have different velocities (\S\ref{sec:pheno_model}). 
We initially allowed the temperatures of the two \texttt{bapec} components to vary freely in the fit. 
We, however, found that the temperature of the component with the smaller velocity broadening, which is responsible for the He-like iron emission, could not be well constrained. 
Thus, by considering a condition that the Fe He$\alpha$ emission should dominate over the Fe Ly$\alpha$ line and emission lines of low-$Z$ ions can be ignored, we representatively fixed the temperature at $kT = 3\,\mathrm{keV}$. 
We note that even if the temperature is varied under the condition, our results are not significantly affected. 
The best-fitting parameters obtained with $C$-stat./d.o.f. = 17025/17298 are listed in Table \ref{tab:bapecfittable}, and the fitting result can be seen in Figure~\ref{fig:rsl_bapec_d_fit}. 
The fit does not improve significantly. 
Thus, as with the photoionization model fit, the additional component is not strongly required.
Figure~\ref{fig:rsl_bapec_d_fit} would suggest the likely reason why the temperature of the narrow \texttt{bapec} component cannot be constrained well. 
While the narrow one is faint in the f line and is likely to be inefficient to reproduce the apparently prominent f and w lines, the broader one seems to be able to more easily account for the line profile. 
Related to this, to reproduce both He$\alpha$ and Ly$\alpha$ lines with comparable fluxes, a high temperature of $11.5^{+5.0}_{-2.0}$~keV would be required for the broader component. 
%

\section{Discussion} \label{sec:4}
\subsection{Fe He$\alpha$ and Ly$\alpha$ Emitting Mechanisms} \label{sec:41_a}
\begin{figure*}[!ht]
 \begin{center}
 \includegraphics[width=1.0\textwidth]{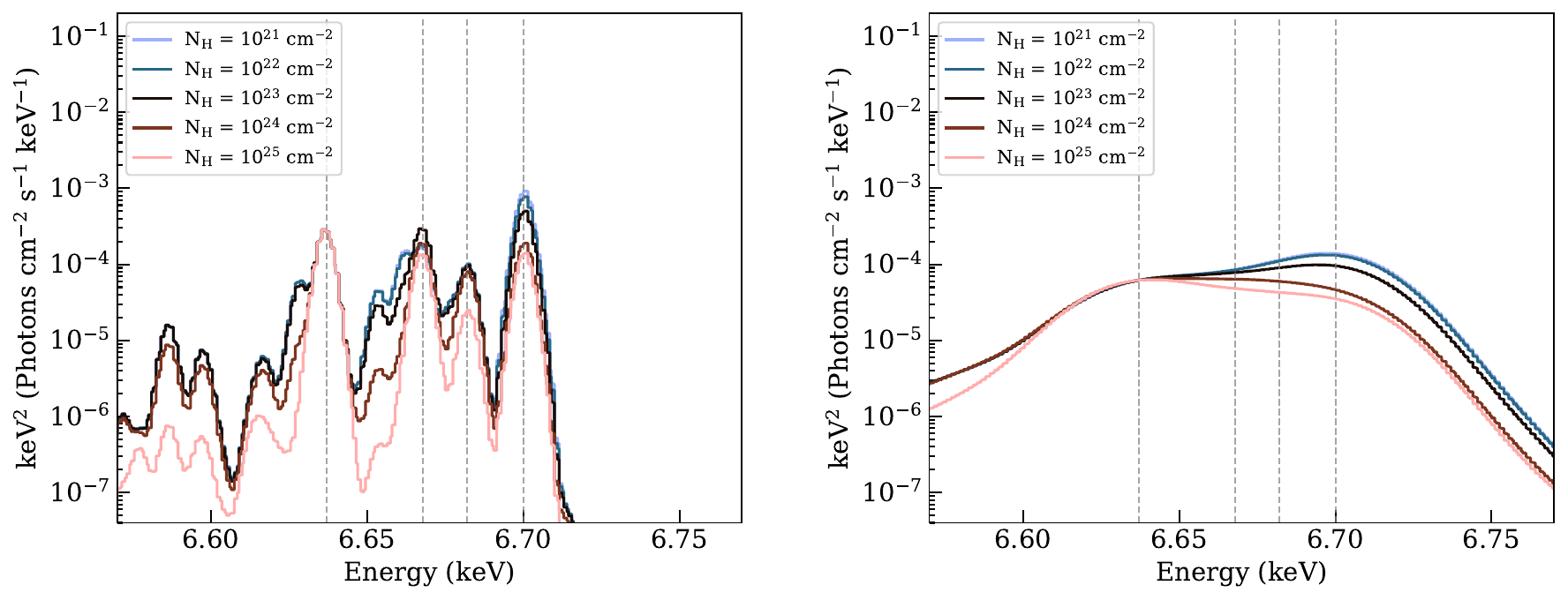}
 \end{center}
 \caption{Spectra in the Fe He$\alpha$ region obtained from our \texttt{pion} model for column densities ranging from $10^{21}$ to $10^{25}$~cm$^{-2}$, normalized to the intensity of the forbidden line at $10^{21}$~cm$^{-2}$. The left panel assumes a velocity broadening of 100~km~s$^{-1}$ for comparison, while the right panel assumes a velocity broadening of 800~km~s$^{-1}$, which was obtained from the fit of the phenomenological model. In both panels, the ionization parameter is set to 3.0 in log units, corresponding to the component explaining Fe He$\alpha$ in the two component photoionization model. The gray dashed lines indicate the Fe He$\alpha$ resonance line, intercombination lines, and forbidden line.
 {Alt text: Two panels showing spectra in the Fe He$\alpha$ region calculated with the pion model for column densities from $10^{21}$ to $10^{25}$~cm$^{-2}$. The left and right panels assume velocity broadenings of 100 and 800~km~s$^{-1}$, respectively. In both panels, the ionization parameter is fixed at 3.0 in log units.}
 }
 \label{fig:pion_spectrum_comparison}
\end{figure*}

\begin{figure*}[!ht]
 \begin{center}
 \includegraphics[width=1.0\textwidth]{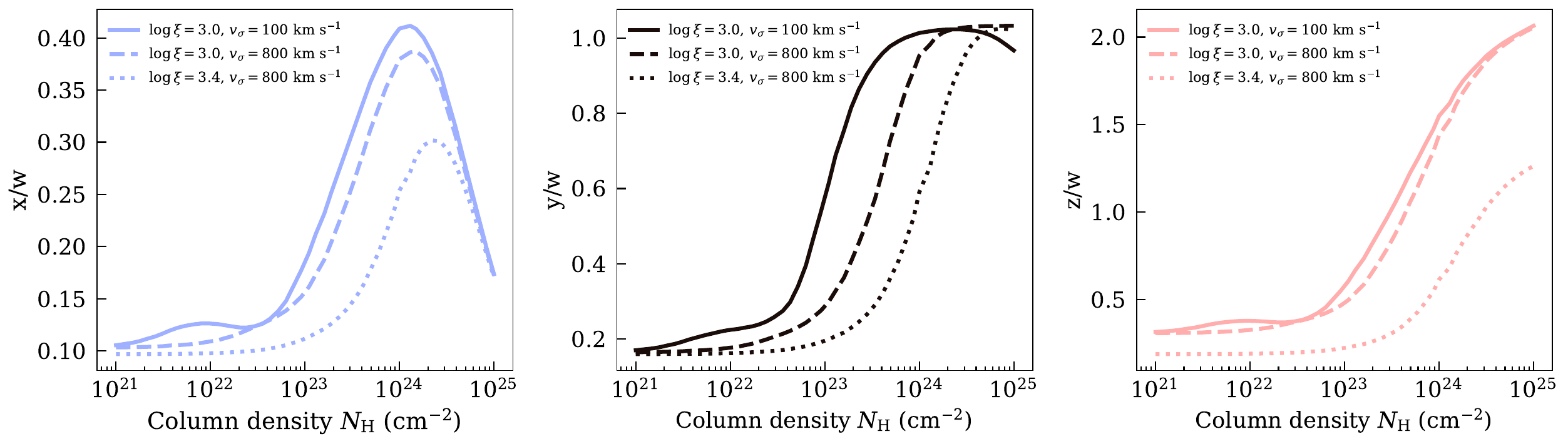}
 \end{center}
 \caption{Line ratios of Fe He$\alpha$ as a function of column density ranging from $10^{21}$ to $10^{25}$~cm$^{-2}$ from our \texttt{pion} model. The left, middle, and right panels show x$/$w, y$/$w, and z$/$w, respectively. In each panel, both the case with a velocity broadening of 800~km~s$^{-1}$, obtained from the fit of the phenomenological model, and the case with 100~km~s$^{-1}$ are shown for comparison. The ionization parameters of 3.4 in log units, obtained from the fit of the single component photoionization model, and 3.0 in log units, corresponding to the component explaining Fe He$\alpha$ in the two component photoionization model, are shown. 
 {Alt text: Three panels showing Fe He$\alpha$ line ratios from the pion model as a function of column density from 10$^{21}$ to 10$^{25}$~cm$^{-2}$. The panels show x over w, y over w, and z over w ratios, comparing velocity broadenings of 100 and 800~km~s$^{-1}$ for ionization parameters of 3.4 and 3.0 in log units.}
 }
 \label{fig:pion_spectrum_line_ratio}
\end{figure*}

To reveal the emitting mechanism of the He$\alpha$ and Ly$\alpha$ lines, after characterizing them using Gaussian functions (\S\ref{sec:pheno_model}), we examined whether photoionization or collisional ionization models can reproduce the observed line profiles. 
In \S\ref{sec:pheno_model}, the Gaussian fits showed that the He$\alpha$ and Ly$\alpha$ lines may be explained by a single ionized plasma, but also potentially by a two-ionized-plasma model. 
Particularly in the latter case, the He$\alpha$ complex appears to be characterized by the f and w lines, which are comparable in flux and stronger than the x$+$y lines.
Motivated by these results, we considered one-zone photoionization and collisional-ionization models, and also two-zone ones. 
Even in the two-zone case, however, the collisional-ionization model generally produces weak forbidden emission and thus cannot reproduce the line-intensity ratios suggested by the Gaussian fitting. 
In contrast, the photoionization models can produce comparably bright forbidden and resonance lines when $N_{\rm H}$ is large. 
The main reason is that the resonance line is suppressed by resonant scattering. 
The spectral change as a function of $N_{\rm H}$, simulated with  
\texttt{pion}, is illustrated in Figure~\ref{fig:pion_spectrum_comparison}, and corresponding quantitative changes in line-intensity ratios are shown in Figure~\ref{fig:pion_spectrum_line_ratio}. 
However, as seen in both figures, the intercombination lines, especially the $y$ line, remain relatively strong in the $N_{\rm H}$ range from $10^{21}$ to $10^{25}$~cm$^{-2}$.
As a result, it seems difficult for the \texttt{pion} model to reproduce the Fe He$\alpha$ line ratios. 
Therefore, even if we replace the one-zone model with the two-zone one, we find no significant improvement in the quality of the fit; this conclusion is also supported by the $C$-statistics (Tables~\ref{tab:pionfittable} and \ref{tab:bapecfittable}). 
Consequently, with the present data, we cannot draw firm conclusions as to whether collisional or photoionization processes dominate, or whether a one-zone or a two-zone configuration is preferred.

In the above discussion, we have implicitly assumed that the ionized plasma resides approximately at the systemic velocity of the host galaxy, such that the forbidden and resonance components are observed near the rest-frame energies. 
If we do not make the assumption, some additional scenarios become possible. 
For example, the emission component from collisionally ionized or photoionized plasma in which the resonance line is strongly enhanced may be present not only at the systemic velocity but also at a redshift corresponding to a recession velocity of $\sim 3000$~km~s$^{-1}$.
A different idea is that slightly energy-shifted absorption by He-like iron could suppress the x+y lines.  
Otherwise, absorption of the x and y lines by ions, including Li-like Fe and He-like Cr, could become significant at high column densities ($N_{\rm H} \gtrsim 10^{23}$~cm$^{-2}$) \citep{Porquet2010, Mehdipour2015, ChakrabortyI2020, ChakrabortyII2020}. 
Exploring these kinds of scenarios in detail, however, would expand the scope of the discussion.
Such a discussion would be fruitful after future XRISM observations provide spectra with higher photon statistics, where a deficit in the x+y energy band may be established with higher significance. 
%

\subsection{Variation of AGN structure with Eddington ratio} \label{sec:43}
Although our spectral fits do not uniquely determine the ionization mechanism for highly ionized iron emission in NGC\,7213, it is still informative to compare NGC\,7213 with M\,81$^\ast$ for probing the Eddington-ratio dependence of AGN structure using the iron lines. 
Among the AGNs observed by XRISM so far, these are the two in which highly ionized iron is detected mainly in emission rather than in absorption \citep{Miller2025}. 
Although it would be fruitful to include AGNs for which highly ionized lines are seen in absorption \citep[e.g.,][]{xrism2025PDS456,Mehdipour2025,Xiang2025}, it makes the discussion complicated. 
This is because we have to consider whether the relevant gas lies along our line of sight to the X-ray source or is distributed around it. 
By focusing on the two AGNs, we can avoid such complications. 
Furthermore, a valuable point for the pair is that the two AGNs have similar central SMBH masses ($\sim 10^8\,M_\odot$) and probably similar viewing angles ($\sim 10$\arcdeg), while their Eddington ratios differ by almost two orders of magnitude. 

The density is a fundamental parameter in both collisional and photoionized plasmas, and we discuss how it depends on the Eddington ratio based on the AGN pair. 
In discussing the dependence, as the ionization mechanism of the iron lines cannot yet be determined uniquely  for both NGC\,7213 and M\,81$^\ast$ \citep{Miller2025}, to keep the discussion simple, we assume that the same mechanism operates in both objects.
Also, for NGC\,7213, we assume the results obtained with the one-zone models for simplicity. 

Based on the EM of $\sim 3\times10^{64}$~cm$^{-3}$ derived from the \texttt{bapec} fits, we estimate the density of the collisionally ionized gas while assuming a representative spherical volume. 
We set the radius of this volume from the line width, adopting Keplerian motion. 
From this simple estimate, we find that the hydrogen density of the gas emitting the ionized iron lines may be higher in NGC\,7213 than in M\,81$^\ast$: $\sim 1\times10^{8}$~cm$^{-3}$ versus $\sim 4\times10^{5}$~cm$^{-3}$. 
For M\,81$^\ast$, \citet{Miller2025} fitted a two-zone collisional model to its Resolve spectrum, and we refer to the dominant collisional component. 
The comparison may suggest that, with the Eddington ratio, the density of highly ionized gas increases. 
This trend may reflect differences in the physical conditions of the circumnuclear gas as a function of the Eddington ratio.

In the collisional-ionization picture, plausible structures for such hot plasma include an optically thin RIAF and/or a wind launched from it.
If a broad neutral iron K$\alpha$ line with an FWHM of $\sim 4500$~km~s$^{-1}$ found in NGC\,7213 \citep{Kammoun2025} traces the inner edge of an optically thick, cold accretion disk, and an RIAF is present only further inside, the observed ionized lines may instead come from a hot wind. 
A similar idea has also been discussed for M~81$^\ast$, for which \citet{Miller2025} argued that the highly ionized iron emission may originate in a hot outflow,  while their estimate of the disk's inner scale is based on UV emission lines. 
If the highly ionized lines indeed come from a hot wind, the redshifted components seen in both NGC\,7213 and M\,81$^\ast$ could indicate a largely equatorial outflow in which emission from the far side is stronger than that from the near side.
Theoretically, winds from hot accretion flows are expected to become stronger as the mass accretion rate increases \citep[e.g.,][]{Yuan2014,Bu2018}.
Therefore, the higher density inferred for the highly ionized iron-line-emitting gas in NGC\,7213 may be understood as a natural consequence of its higher accretion rate (higher $\lambda_{\rm Edd}$) compared to M\,81$^\ast$.

In the photoionization picture, we assumed a gas density of $10^{8}$~cm$^{-3}$ for NGC\,7213 and showed that this value is consistent with the fitted ionization parameter.
For M~81$^\ast$, the ionization parameter of $\log\xi=3.16$ \citep[Table~2 of][]{Miller2025} similarly implies a density of $\sim10^{4}$~cm$^{-3}$, again lower than that of NGC\,7213.

A possible location for the photoionized gas is, for instance, above the disk, including the BLR and the disk atmosphere.
This density is too low for the accretion disk.
Although typical lower limits on BLR cloud densities are $\sim10^{8}$~cm$^{-3}$, consistent with our assumed value, a BLR origin may not be so favored as discussed below. 
In general, the characteristic BLR radius is expected to increase with luminosity \citep[e.g.,][]{Kaspi2000,Bentz2013}, so the higher-luminosity source would have a larger BLR scale.
This trend does not match the highly ionized iron lines in NGC\,7213 and M\,81$^\ast$, where the line widths suggest that the ionized iron emission in NGC\,7213 can originate from smaller radii than in M\,81$^\ast$.
Instead, the disk-atmosphere scenario might be more plausible \citep{Begelman1983,Tomaru2019,Tomaru2023,Waters2021}. 
At higher Eddington ratios, the cold, optically thick disk may extend further inward \citep[e.g.,][]{Yuan2004}, so that the disk atmosphere at small radii may be denser in NGC\,7213 than in M~81$^\ast$.
In this picture, a higher luminosity does not necessarily shift the 
Fe H$\alpha$ and Ly$\alpha$ emitting region to larger radii.

In summary, under both collisional and photoionization scenarios, our results seem to favor a picture in which the density of the gas emitting highly ionized iron lines decreases toward lower Eddington ratios. 
This trend may be natural given that the lower Eddington ratio reflects a lower mass accretion rate onto the nuclear region.

\section{Conclusions} \label{sec:conc}
Based on XRISM and NuSTAR spectra obtained for NGC\,7213 at $\lambda_{\rm Edd} \sim 10^{-3}$, we investigated whether either a photoionization or a collisional ionization model can reproduce Fe He$\alpha$ and Ly$\alpha$ emission lines in the Resolve spectrum. 
First, after determining the shape of the continuum emission with the help of NuSTAR, we phenomenologically characterized Fe He$\alpha$ and Fe Ly$\alpha$ in the Resolve spectrum with Gaussian functions  (\S\ref{sec:pheno_model}). 
The result showed that the line width of the Fe Ly$\alpha$ lines ($\approx 2600$~km~s$^{-1}$) may be larger than that of the Fe He$\alpha$ lines ($\approx 800$~km~s$^{-1}$), although the difference is not significant. 
Considering the possible difference in line widths, we fitted not only one-zone ionized plasma models, but also two-zone models (\S\ref{sec:photo_model} and \S\ref{sec:col_model}). 
Results are that the additional ionized component is not significantly required, and also the current data do not uniquely determine whether photoionization or collisional ionization dominates. 
It also appears challenging to reproduce the He$\alpha$ profile with either the collisional ionization model or photoionization one adopted in this study, if the line profile, suggested by the Gaussian fits, indicates suppressed x+y lines relative to the w and z lines.
Finally, we compared NGC\,7213 with M~81$^\ast$, another XRISM target at a lower Eddington ratio of $\lambda_{\rm Edd}\sim10^{-5}$.  
Under the assumption that the same ionization mechanism operates in both objects, we find that the density of the highly ionized iron-line-emitting gas may be higher in NGC\,7213 than in M\,81$^\ast$, regardless of the assumed ionizing mechanism. 
This might reflect that lower-$\lambda_{\rm Edd}$ AGNs have lower-density gas. 
Expanding observations to a broader range of Eddington ratios and larger samples will further clarify the relationship between the physical properties of the ionized gas, such as density, and the Eddington ratio across a wide parameter space.

\section*{Funding}
T.K., R.T., K.O., H.N., and H.M. are supported by JSPS KAKENHI grant numbers 23K13153$/$24K00673, 24KJ0152, 22H00128$/$22K18277, 23K20239$/$24K00672$/$25H00660, and 23H00128, respectively.
S.B. acknowledges funding from PRIN MUR 2022 SEAWIND 2022Y2T94C, supported by European Union $-$ Next Generation EU, Mission 4 Component 1 CUP C53D23001330006.
C.R. acknowledges support from SNSF Consolidator grant F01$-$13252, Fondecyt Regular grant 1230345, ANID BASAL project FB210003 and the China-Chile joint research fund. This work was supported by JSPS 
Core-to-Core Program, (grant number:JPJSCCA20220002). 

\begin{ack}
The XRISM observation provided new findings and further motivated our interest. 
This work was made possible by the dedicated efforts of the XRISM team members, engineers, and operations team, and we thank them for their efforts.
The NuSTAR observation enhanced the scientific value of the XRISM data, and we also thank all members of the NuSTAR mission.
Additionally, we thank the referee for their helpful comments and Yoshiyuki Inoue for valuable discussions.

\end{ack}

\bibliography{main_pasj}{}

@ARTICLE{cash1979,
       author = {{Cash}, W.},
        title = "{Parameter estimation in astronomy through application of the likelihood ratio.}",
      journal = {\apj},
         year = 1979,
        month = mar,
       volume = {228},
        pages = {939-947},
          doi = {10.1086/156922},
       adsurl = {https://ui.adsabs.harvard.edu/abs/1979ApJ...228..939C}
}

@INPROCEEDINGS{Arnaud1996,
       author = {{Arnaud}, K.~A.},
        title = "{XSPEC: The First Ten Years}",
    booktitle = {Astronomical Data Analysis Software and Systems V},
         year = 1996,
       editor = {{Jacoby}, George H. and {Barnes}, Jeannette},
       series = {Astronomical Society of the Pacific Conference Series},
       volume = {101},
        month = jan,
        pages = {17},
       adsurl = {https://ui.adsabs.harvard.edu/abs/1996ASPC..101...17A}
}

@ARTICLE{Mehdipour2016,
       author = {{Mehdipour}, M. and {Kaastra}, J.~S. and {Kallman}, T.},
        title = "{Systematic comparison of photoionised plasma codes with application to spectroscopic studies of AGN in X-rays}",
      journal = {\aap},
         year = 2016,
        month = dec,
       volume = {596},
          eid = {A65},
        pages = {A65},
          doi = {10.1051/0004-6361/201628721},
archivePrefix = {arXiv},
       eprint = {1610.03080},
 primaryClass = {astro-ph.HE},
       adsurl = {https://ui.adsabs.harvard.edu/abs/2016A&A...596A..65M}
}

@INPROCEEDINGS{Kaastra1996,
       author = {{Kaastra}, J.~S. and {Mewe}, R. and {Nieuwenhuijzen}, H.},
        title = "{SPEX: a new code for spectral analysis of X \& UV spectra.}",
    booktitle = {UV and X-ray Spectroscopy of Astrophysical and Laboratory Plasmas},
         year = 1996,
       editor = {{Yamashita}, K. and {Watanabe}, T.},
        month = jan,
        pages = {411-414},
       adsurl = {https://ui.adsabs.harvard.edu/abs/1996uxsa.conf..411K}
}

@ARTICLE{Noda2025,
       author = {{Noda}, Hirofumi and {Mori}, Koji and {Tomida}, Hiroshi and {Nakajima}, Hiroshi and {Tanaka}, Takaaki and {Murakami}, Hiroshi and {Uchida}, Hiroyuki and {Suzuki}, Hiromasa and {Kobayashi}, Shogo Benjamin and {Yoneyama}, Tomokage and {Hagino}, Kouichi and {Nobukawa}, Kumiko and {Uchiyama}, Hideki and {Nobukawa}, Masayoshi and {Matsumoto}, Hironori and {Tsuru}, Takeshi Go and {Yamauchi}, Makoto and {Hatsukade}, Isamu and {Odaka}, Hirokazu and {Kohmura}, Takayoshi and {Yamaoka}, Kazutaka and {Yoshida}, Tessei and {Kanemaru}, Yoshiaki and {Hiraga}, Junko and {Dotani}, Tadayasu and {Ozaki}, Masanobu and {Tsunemi}, Hiroshi and {Sato}, Jin and {Takaki}, Toshiyuki and {Terada}, Yuta and {Miyazaki}, Keitaro and {Kusunoki}, Kohei and {Otsuka}, Yoshinori and {Yokosu}, Haruhiko and {Yonemaru}, Wakana and {Ichikawa}, Kazuhiro and {Nakano}, Hanako and {Takemoto}, Reo and {Matsushima}, Tsukasa and {Urase}, Reika and {Kurashima}, Jun and {Fuchi}, Kotomi and {Hayakawa}, Kaito and {Fukuda}, Masahiro and {Kamei}, Takamitsu and {Asahina}, Yoh and {Inoue}, Shun and {Amano}, Yuki and {Aoki}, Yuma and {Ito}, Yamato and {Kamatani}, Tomoya and {Takayama}, Kouta and {Sako}, Takashi and {Yoshimoto}, Marina and {Shima}, Kohei and {Higuchi}, Mayu and {Ninoyu}, Kaito and {Aoki}, Daiki and {Tsunomachi}, Shun and {Hayashida}, Kiyoshi},
        title = "{Soft X-ray Imager of the Xtend system on board XRISM}",
      journal = {\pasj},
         year = 2025,
        month = mar,
          doi = {10.1093/pasj/psaf011},
archivePrefix = {arXiv},
       eprint = {2502.08030},
 primaryClass = {astro-ph.IM},
       adsurl = {https://ui.adsabs.harvard.edu/abs/2025PASJ..tmp...16N}
}

@article{Ishisaki2025,
author = {Yoshitaka Ishisaki and Richard L. Kelley and Hisamitsu Awaki and Jesus C. Balleza and Kim R. Barnstable and Thomas G. Bialas and Rozenn Boissay-Malaquin and Gregory V. Brown and Edgar R. Canavan and Renata S. Cumbee and Timothy M. Carnahan and Meng P. Chiao and Brian J. Comber and Elisa Costantini and Jan-Willem den Herder and Johannes Dercksen and Cor P. de Vries and Michael J. DiPirro and Megan E. Eckart and Yuichiro Ezoe and Carlo Ferrigno and Ryuichi Fujimoto and Nathalie Gorter and Steven M. Graham and Martin Grim and Leslie S. Hartz and Ryota Hayakawa and Takayuki Hayashi and Natalie Hell and Akio Hoshino and Yuto Ichinohe and Manabu Ishida and Kumi Ishikawa and Bryan L. James and Steven J. Kenyon and Caroline A. Kilbourne and Mark O. Kimball and Shunji Kitamoto and Maurice A. Leutenegger and Yoshitomo Maeda and Dan McCammon and Joseph J. Miko and Misaki Mizumoto and Hirofumi Noda and Takashi Okajima and Atsushi Okamoto and Stephane Paltani and Frederick S. Porter and Kosuke Sato and Toshiki Sato and Makoto Sawada and Keisuke Shinozaki and Russell Shipman and Peter J. Shirron and Gary A. Sneiderman and Yang Soong and Richard Szymkiewicz and Andrew E. Szymkowiak and Yoh Takei and Keisuke Tamura and Masahiro Tsujimoto and Yuusuke Uchida and Stephen Wasserzug and Michael C. Witthoeft and Rob Wolfs and Shinya Yamada and Susumu Yasuda},
title = {{Resolve instrument onboard XRISM: design, integration, and instrument test results}},
volume = {11},
journal = {Journal of Astronomical Telescopes, Instruments, and Systems},
number = {4},
publisher = {SPIE},
pages = {042023},
year = {2025},
doi = {10.1117/1.JATIS.11.4.042023},
URL = {https://doi.org/10.1117/1.JATIS.11.4.042023}
}

@ARTICLE{Phillips1979,
       author = {{Phillips}, M.~M.},
        title = "{Optical Spectrophotometry of the Suspected X-Ray Galaxies NGC 6221 and NGC 7213}",
      journal = {\apjl},
         year = 1979,
        month = feb,
       volume = {227},
        pages = {L121},
          doi = {10.1086/182881},
       adsurl = {https://ui.adsabs.harvard.edu/abs/1979ApJ...227L.121P}
}

@ARTICLE{HI4PI2016,
       author = {{HI4PI Collaboration} and {Ben Bekhti}, N. and {Fl{\"o}er}, L. and {Keller}, R. and {Kerp}, J. and {Lenz}, D. and {Winkel}, B. and {Bailin}, J. and {Calabretta}, M.~R. and {Dedes}, L. and {Ford}, H.~A. and {Gibson}, B.~K. and {Haud}, U. and {Janowiecki}, S. and {Kalberla}, P.~M.~W. and {Lockman}, F.~J. and {McClure-Griffiths}, N.~M. and {Murphy}, T. and {Nakanishi}, H. and {Pisano}, D.~J. and {Staveley-Smith}, L.},
        title = "{HI4PI: A full-sky H I survey based on EBHIS and GASS}",
      journal = {\aap},
         year = 2016,
        month = oct,
       volume = {594},
          eid = {A116},
        pages = {A116},
          doi = {10.1051/0004-6361/201629178},
archivePrefix = {arXiv},
       eprint = {1610.06175},
 primaryClass = {astro-ph.GA},
       adsurl = {https://ui.adsabs.harvard.edu/abs/2016A&A...594A.116H}
}

@ARTICLE{Lobban2010,
       author = {{Lobban}, A.~P. and {Reeves}, J.~N. and {Porquet}, D. and {Braito}, V. and {Markowitz}, A. and {Miller}, L. and {Turner}, T.~J.},
        title = "{Evidence for a truncated accretion disc in the low-luminosity Seyfert galaxy, NGC 7213?}",
      journal = {\mnras},
         year = 2010,
        month = oct,
       volume = {408},
       number = {1},
        pages = {551-564},
          doi = {10.1111/j.1365-2966.2010.17143.x},
archivePrefix = {arXiv},
       eprint = {1006.1318},
 primaryClass = {astro-ph.HE},
       adsurl = {https://ui.adsabs.harvard.edu/abs/2010MNRAS.408..551L}
}

@ARTICLE{Emmanoulopoulos2013,
       author = {{Emmanoulopoulos}, D. and {Papadakis}, I.~E. and {Nicastro}, F. and {McHardy}, I.~M.},
        title = "{X-ray spectral analysis of the low-luminosity active galactic nucleus NGC 7213 using long XMM-Newton observations}",
      journal = {\mnras},
         year = 2013,
        month = mar,
       volume = {429},
       number = {4},
        pages = {3439-3448},
          doi = {10.1093/mnras/sts610},
archivePrefix = {arXiv},
       eprint = {1212.2853},
 primaryClass = {astro-ph.HE},
       adsurl = {https://ui.adsabs.harvard.edu/abs/2013MNRAS.429.3439E}
}

@ARTICLE{Woo2002,
       author = {{Woo}, Jong-Hak and {Urry}, C. Megan},
        title = "{Active Galactic Nucleus Black Hole Masses and Bolometric Luminosities}",
      journal = {\apj},
         year = 2002,
        month = nov,
       volume = {579},
       number = {2},
        pages = {530-544},
          doi = {10.1086/342878},
archivePrefix = {arXiv},
       eprint = {astro-ph/0207249},
 primaryClass = {astro-ph},
       adsurl = {https://ui.adsabs.harvard.edu/abs/2002ApJ...579..530W}
}

@ARTICLE{Miller2025,
       author = {{Miller}, Jon M. and {Behar}, Ehud and {Awaki}, Hisamitsu and {Hornschemeier}, Ann and {Bluem}, Jesse and {Gallo}, Luigi and {Kobayashi}, Shogo B. and {Mushotzky}, Richard and {Ohno}, Masanori and {Petre}, Robert and {Sato}, Kosuke and {Terashima}, Yuichi and {Yukita}, Mihoko},
        title = "{XRISM Reveals a Remnant Torus in the Low-luminosity AGN M81*}",
      journal = {\apjl},
         year = 2025,
        month = jun,
       volume = {985},
       number = {2},
          eid = {L41},
        pages = {L41},
          doi = {10.3847/2041-8213/add262},
archivePrefix = {arXiv},
       eprint = {2505.13730},
 primaryClass = {astro-ph.HE},
       adsurl = {https://ui.adsabs.harvard.edu/abs/2025ApJ...985L..41M}
}

@ARTICLE{Begelman1983,
       author = {{Begelman}, M.~C. and {McKee}, C.~F. and {Shields}, G.~A.},
        title = "{Compton heated winds and coronae above accretion disks. I. Dynamics.}",
      journal = {\apj},
         year = 1983,
        month = aug,
       volume = {271},
        pages = {70-88},
          doi = {10.1086/161178},
       adsurl = {https://ui.adsabs.harvard.edu/abs/1983ApJ...271...70B}
}

@ARTICLE{Tomaru2023,
       author = {{Tomaru}, Ryota and {Done}, Chris and {Odaka}, Hirokazu and {Tanimoto}, Atsushi},
        title = "{A different view of wind in X-ray binaries: the accretion disc corona source 2S 0921-630}",
      journal = {\mnras},
         year = 2023,
        month = aug,
       volume = {523},
       number = {3},
        pages = {3441-3449},
          doi = {10.1093/mnras/stad1637},
archivePrefix = {arXiv},
       eprint = {2302.12638},
 primaryClass = {astro-ph.HE},
       adsurl = {https://ui.adsabs.harvard.edu/abs/2023MNRAS.523.3441T}
}

@ARTICLE{Duras2020,
       author = {{Duras}, F. and {Bongiorno}, A. and {Ricci}, F. and {Piconcelli}, E. and {Shankar}, F. and {Lusso}, E. and {Bianchi}, S. and {Fiore}, F. and {Maiolino}, R. and {Marconi}, A. and {Onori}, F. and {Sani}, E. and {Schneider}, R. and {Vignali}, C. and {La Franca}, F.},
        title = "{Universal bolometric corrections for active galactic nuclei over seven luminosity decades}",
      journal = {\aap},
         year = 2020,
        month = apr,
       volume = {636},
          eid = {A73},
        pages = {A73},
          doi = {10.1051/0004-6361/201936817},
archivePrefix = {arXiv},
       eprint = {2001.09984},
 primaryClass = {astro-ph.GA},
       adsurl = {https://ui.adsabs.harvard.edu/abs/2020A&A...636A..73D}
}

@ARTICLE{Yan2018,
       author = {{Yan}, Zhen and {Xie}, Fu-Guo},
        title = "{A decades-long fast-rise-exponential-decay flare in low-luminosity AGN NGC 7213}",
      journal = {\mnras},
         year = 2018,
        month = mar,
       volume = {475},
       number = {1},
        pages = {1190-1197},
          doi = {10.1093/mnras/stx3259},
archivePrefix = {arXiv},
       eprint = {1712.05272},
 primaryClass = {astro-ph.HE},
       adsurl = {https://ui.adsabs.harvard.edu/abs/2018MNRAS.475.1190Y}
}

@ARTICLE{Bianchi2008,
       author = {{Bianchi}, Stefano and {La Franca}, Fabio and {Matt}, Giorgio and {Guainazzi}, Matteo and {Jimenez Bail{\'o}n}, Elena and {Longinotti}, Anna Lia and {Nicastro}, Fabrizio and {Pentericci}, Laura},
        title = "{A broad-line region origin for the iron K{\ensuremath{\alpha}} line in NGC 7213}",
      journal = {\mnras},
         year = 2008,
        month = sep,
       volume = {389},
       number = {1},
        pages = {L52-L56},
          doi = {10.1111/j.1745-3933.2008.00521.x},
archivePrefix = {arXiv},
       eprint = {0806.3876},
 primaryClass = {astro-ph},
       adsurl = {https://ui.adsabs.harvard.edu/abs/2008MNRAS.389L..52B}
}

@ARTICLE{Netzer2015,
       author = {{Netzer}, Hagai},
        title = "{Revisiting the Unified Model of Active Galactic Nuclei}",
      journal = {\araa},
         year = 2015,
        month = aug,
       volume = {53},
        pages = {365-408},
          doi = {10.1146/annurev-astro-082214-122302},
archivePrefix = {arXiv},
       eprint = {1505.00811},
 primaryClass = {astro-ph.GA},
       adsurl = {https://ui.adsabs.harvard.edu/abs/2015ARA&A..53..365N}
}

@ARTICLE{Giustini2019,
       author = {{Giustini}, Margherita and {Proga}, Daniel},
        title = "{A global view of the inner accretion and ejection flow around super massive black holes. Radiation-driven accretion disk winds in a physical context}",
      journal = {\aap},
         year = 2019,
        month = oct,
       volume = {630},
          eid = {A94},
        pages = {A94},
          doi = {10.1051/0004-6361/201833810},
archivePrefix = {arXiv},
       eprint = {1904.07341},
 primaryClass = {astro-ph.GA},
       adsurl = {https://ui.adsabs.harvard.edu/abs/2019A&A...630A..94G}
}

@ARTICLE{Yuan2014,
       author = {{Yuan}, Feng and {Narayan}, Ramesh},
        title = "{Hot Accretion Flows Around Black Holes}",
      journal = {\araa},
         year = 2014,
        month = aug,
       volume = {52},
        pages = {529-588},
          doi = {10.1146/annurev-astro-082812-141003},
archivePrefix = {arXiv},
       eprint = {1401.0586},
 primaryClass = {astro-ph.HE},
       adsurl = {https://ui.adsabs.harvard.edu/abs/2014ARA&A..52..529Y}
}

@ARTICLE{Ho2008,
       author = {{Ho}, L.~C.},
        title = "{Nuclear activity in nearby galaxies.}",
      journal = {\araa},
         year = 2008,
        month = sep,
       volume = {46},
        pages = {475-539},
          doi = {10.1146/annurev.astro.45.051806.110546},
archivePrefix = {arXiv},
       eprint = {0803.2268},
 primaryClass = {astro-ph},
       adsurl = {https://ui.adsabs.harvard.edu/abs/2008ARA&A..46..475H}
}

@BOOK{Peterson1997,
       author = {{Peterson}, Bradley M.},
        title = "{An Introduction to Active Galactic Nuclei}",
         year = 1997,
       adsurl = {https://ui.adsabs.harvard.edu/abs/1997iagn.book.....P}
}

@ARTICLE{Xiang2025,
       author = {{Xiang}, Xin and {Miller}, Jon M. and {Behar}, Ehud and {Boissay-Malaquin}, Rozenn and {Brenneman}, Laura and {Buhariwalla}, Margaret and {Byun}, Doyee and {Done}, Chris and {Gallo}, Luigi and {Gerolymatou}, Dimitra and {Hagen}, Scott and {Kaastra}, Jelle and {Paltani}, Stephane and {Porter}, Frederick S. and {Mushotzky}, Richard and {Noda}, Hirofumi and {Mehdipour}, Missagh and {Minezaki}, Takeo and {Tashiro}, Makoto and {Zoghbi}, Abderahmen},
        title = "{XRISM Spectroscopy of Accretion-driven Wind Feedback in NGC 4151}",
      journal = {\apjl},
         year = 2025,
        month = aug,
       volume = {988},
       number = {2},
          eid = {L54},
        pages = {L54},
          doi = {10.3847/2041-8213/adee9b},
archivePrefix = {arXiv},
       eprint = {2507.09210},
 primaryClass = {astro-ph.HE},
       adsurl = {https://ui.adsabs.harvard.edu/abs/2025ApJ...988L..54X}
}

@misc{Kammoun2025,
      title={XRISM/Resolve reveals the complex iron structure of NGC 7213: Evidence for radial stratification between inner disk and broad-line region}, 
      author={E. Kammoun and T. Kawamuro and K. Murakami and S. Bianchi and F. Nicastro and A. Luminari and E. Aydi and M. Eracleous and O. K. Adegoke and E. Bertola and P. G. Boorman and V. Braito and G. Bruni and A. Comastri and P. Condò and M. Dadina and T. Enoto and J. A. García and V. E. Gianolli and F. A. Harrison and G. Lanzuisi and M. Laurenti and A. Marinucci and G. Mastroserio and H. Matsumoto and G. Matt and G. Matzeu and R. Middei and E. Nardini and H. Noda and H. Odaka and S. Ogawa and F. Panessa and E. Piconcelli and C. Pinto and J. M. Piotrowska and G. Ponti and C. Ricci and R. Ricci and R. Serafinelli and F. Shi and D. Stern and A. Tanimoto and Y. Terashima and R. Tomaru and F. Tombesi and A. Tortosa and Y. Ueda and F. Ursini and C. Vignali and S. Yamada and S. Yamada},
      year={2025},
      eprint={2510.24971},
      archivePrefix={arXiv},
      primaryClass={astro-ph.HE},
      url={https://arxiv.org/abs/2510.24971}, 
}

@ARTICLE{xrism2025PDS456,
       author = {{Xrism Collaboration} and {Audard}, Marc and {Awaki}, Hisamitsu and {Ballhausen}, Ralf and {Bamba}, Aya and {Behar}, Ehud and {Boissay-Malaquin}, Rozenn and {Brenneman}, Laura and {Brown}, Gregory V. and {Corrales}, Lia and {Costantini}, Elisa and {Cumbee}, Renata and {Trigo}, Mar{\'\i}a D{\'\i}az and {Done}, Chris and {Dotani}, Tadayasu and {Ebisawa}, Ken and {Eckart}, Megan and {Eckert}, Dominique and {Enoto}, Teruaki and {Eguchi}, Satoshi and {Ezoe}, Yuichiro and {Foster}, Adam and {Fujimoto}, Ryuichi and {Fujita}, Yutaka and {Fukazawa}, Yasushi and {Fukushima}, Kotaro and {Furuzawa}, Akihiro and {Gallo}, Luigi and {Garc{\'\i}a}, Javier A. and {Gu}, Liyi and {Guainazzi}, Matteo and {Hagino}, Kouichi and {Hamaguchi}, Kenji and {Hatsukade}, Isamu and {Hayashi}, Katsuhiro and {Hayashi}, Takayuki and {Hell}, Natalie and {Hodges-Kluck}, Edmund and {Hornschemeier}, Ann and {Ichinohe}, Yuto and {Ishida}, Manabu and {Ishikawa}, Kumi and {Ishisaki}, Yoshitaka and {Kaastra}, Jelle and {Kallman}, Timothy and {Kara}, Erin and {Katsuda}, Satoru and {Kanemaru}, Yoshiaki and {Kelley}, Richard and {Kilbourne}, Caroline and {Kitamoto}, Shunji and {Kobayashi}, Shogo and {Kohmura}, Takayoshi and {Kubota}, Aya and {Leutenegger}, Maurice and {Loewenstein}, Michael and {Maeda}, Yoshitomo and {Markevitch}, Maxim and {Matsumoto}, Hironori and {Matsushita}, Kyoko and {McCammon}, Dan and {McNamara}, Brian and {Mernier}, Fran{\c{c}}ois and {Miller}, Eric D. and {Miller}, Jon M. and {Mitsuishi}, Ikuyuki and {Mizumoto}, Misaki and {Mizuno}, Tsunefumi and {Mori}, Koji and {Mukai}, Koji and {Murakami}, Hiroshi and {Mushotzky}, Richard and {Nakajima}, Hiroshi and {Nakazawa}, Kazuhiro and {Ness}, Jan-Uwe and {Nobukawa}, Kumiko and {Nobukawa}, Masayoshi and {Noda}, Hirofumi and {Odaka}, Hirokazu and {Ogawa}, Shoji and {Ogorzalek}, Anna and {Okajima}, Takashi and {Ota}, Naomi and {Paltani}, Stephane and {Petre}, Robert and {Plucinsky}, Paul and {Porter}, Frederick Scott and {Pottschmidt}, Katja and {Sato}, Kosuke and {Sato}, Toshiki and {Sawada}, Makoto and {Seta}, Hiromi and {Shidatsu}, Megumi and {Simionescu}, Aurora and {Smith}, Randall and {Suzuki}, Hiromasa and {Szymkowiak}, Andrew and {Takahashi}, Hiromitsu and {Takeo}, Mai and {Tamagawa}, Toru and {Tamura}, Keisuke and {Tanaka}, Takaaki and {Tanimoto}, Atsushi and {Tashiro}, Makoto and {Terada}, Yukikatsu and {Terashima}, Yuichi and {Tsuboi}, Yohko and {Tsujimoto}, Masahiro and {Tsunemi}, Hiroshi and {Tsuru}, Takeshi G. and {Uchida}, Hiroyuki and {Uchida}, Nagomi and {Uchida}, Yuusuke and {Uchiyama}, Hideki and {Ueda}, Yoshihiro and {Uno}, Shinichiro and {Vink}, Jacco and {Watanabe}, Shin and {Williams}, Brian J. and {Yamada}, Satoshi and {Yamada}, Shinya and {Yamaguchi}, Hiroya and {Yamaoka}, Kazutaka and {Yamasaki}, Noriko and {Yamauchi}, Makoto and {Yamauchi}, Shigeo and {Yaqoob}, Tahir and {Yoneyama}, Tomokage and {Yoshida}, Tessei and {Yukita}, Mihoko and {Zhuravleva}, Irina and {Braito}, Valentina and {Cond{\`o}}, Pierpaolo and {Fukumura}, Keigo and {Gonzalez}, Adam and {Luminari}, Alfredo and {Miyamoto}, Aiko and {Mizukawa}, Ryuki and {Reeves}, James and {Sato}, Riki and {Tombesi}, Francesco and {Xu}, Yerong},
        title = "{Structured ionized winds shooting out from a quasar at relativistic speeds}",
      journal = {\nat},
         year = 2025,
        month = may,
       volume = {641},
       number = {8065},
        pages = {1132-1136},
          doi = {10.1038/s41586-025-08968-2},
archivePrefix = {arXiv},
       eprint = {2505.09171},
 primaryClass = {astro-ph.HE},
       adsurl = {https://ui.adsabs.harvard.edu/abs/2025Natur.641.1132X}
}

@article{Takeuchi2013,
   title={Clumpy Outflows from Supercritical Accretion Flow},
   volume={65},
   ISSN={0004-6264},
   url={http://dx.doi.org/10.1093/pasj/65.4.88},
   DOI={10.1093/pasj/65.4.88},
   number={4},
   journal={Publications of the Astronomical Society of Japan},
   publisher={Oxford University Press (OUP)},
   author={Takeuchi, Shun and Ohsuga, Ken and Mineshige, Shin},
   year={2013},
   month=aug }

@article{Young2007,
title = "High-resolution X-ray spectroscopy of a low-luminosity active galactic nucleus: The structure and dynamics of M81",
author = "Young, \{A. J.\} and Nowak, \{M. A.\} and S. Markoff and Marshall, \{H. L.\} and Caneares, \{C. R.\}",
year = "2007",
month = nov,
day = "10",
doi = "10.1086/521778",
language = "English",
volume = "669",
pages = "830--840",
journal = "Astrophysical Journal",
issn = "0004-637X",
number = "2",
}

@article{Shi2022,
doi = {10.3847/1538-4357/ac4789},
url = {https://doi.org/10.3847/1538-4357/ac4789},
year = {2022},
month = {feb},
publisher = {The American Astronomical Society},
volume = {926},
number = {2},
pages = {209},
author = {Shi, Fangzheng and Zhu, Bocheng and Li, Zhiyuan and Yuan, Feng},
title = {Evidence for A Hot Wind from High-resolution X-Ray Spectroscopic Observation of the Low-luminosity Active Galactic Nucleus in NGC 7213},
journal = {The Astrophysical Journal},
}

@ARTICLE{ChakrabortyI2020,
       author = {{Chakraborty}, P. and {Ferland}, G.~J. and {Chatzikos}, M. and {Guzm{\'a}n}, F. and {Su}, Y.},
        title = "{X-Ray Spectroscopy in the Microcalorimeter Era. I. Effects of Fe XXIV Resonant Auger Destruction on Fe XXV K{\ensuremath{\alpha}} Spectra}",
      journal = {\apj},
         year = 2020,
        month = sep,
       volume = {901},
       number = {1},
          eid = {68},
        pages = {68},
          doi = {10.3847/1538-4357/abaaab},
archivePrefix = {arXiv},
       eprint = {2007.15565},
 primaryClass = {astro-ph.HE},
       adsurl = {https://ui.adsabs.harvard.edu/abs/2020ApJ...901...68C}
}

@ARTICLE{ChakrabortyII2020,
       author = {{Chakraborty}, P. and {Ferland}, G.~J. and {Chatzikos}, M. and {Guzm{\'a}n}, F. and {Su}, Y.},
        title = "{X-Ray Spectroscopy in the Microcalorimeter Era. II. A New Diagnostic on Column Density from the Case A to B Transition in H- and He-like Iron}",
      journal = {\apj},
         year = 2020,
        month = sep,
       volume = {901},
       number = {1},
          eid = {69},
        pages = {69},
          doi = {10.3847/1538-4357/abaaac},
archivePrefix = {arXiv},
       eprint = {2007.16106},
 primaryClass = {astro-ph.HE},
       adsurl = {https://ui.adsabs.harvard.edu/abs/2020ApJ...901...69C}
}

@ARTICLE{Mehdipour2015,
       author = {{Mehdipour}, M. and {Kaastra}, J.~S. and {Raassen}, A.~J.~J.},
        title = "{Line absorption of He-like triplet lines by Li-like ions. Caveats of using line ratios of triplets for plasma diagnostics}",
      journal = {\aap},
         year = 2015,
        month = jul,
       volume = {579},
          eid = {A87},
        pages = {A87},
          doi = {10.1051/0004-6361/201526324},
archivePrefix = {arXiv},
       eprint = {1505.06034},
 primaryClass = {astro-ph.HE},
       adsurl = {https://ui.adsabs.harvard.edu/abs/2015A&A...579A..87M}
}

@article{Porquet2010,
   title={He-like Ions as Practical Astrophysical Plasma Diagnostics: From Stellar Coronae to Active Galactic Nuclei},
   volume={157},
   ISSN={1572-9672},
   url={http://dx.doi.org/10.1007/s11214-010-9731-2},
   DOI={10.1007/s11214-010-9731-2},
   number={1–4},
   journal={Space Science Reviews},
   publisher={Springer Science and Business Media LLC},
   author={Porquet, D. and Dubau, J. and Grosso, N.},
   year={2010},
   month=dec, pages={103–134} 
}

@ARTICLE{Bu2018,
       author = {{Bu}, De-Fu and {Gan}, Zhao-Ming},
        title = "{On the wind production from hot accretion flows with different accretion rates}",
      journal = {\mnras},
         year = 2018,
        month = feb,
       volume = {474},
       number = {1},
        pages = {1206-1213},
          doi = {10.1093/mnras/stx2894},
archivePrefix = {arXiv},
       eprint = {1711.02238},
 primaryClass = {astro-ph.HE},
       adsurl = {https://ui.adsabs.harvard.edu/abs/2018MNRAS.474.1206B}
}

@ARTICLE{Yuan2004,
       author = {{Yuan}, Feng and {Narayan}, Ramesh},
        title = "{On the Nature of X-Ray-Bright, Optically Normal Galaxies}",
      journal = {\apj},
         year = 2004,
        month = sep,
       volume = {612},
       number = {2},
        pages = {724-728},
          doi = {10.1086/422802},
archivePrefix = {arXiv},
       eprint = {astro-ph/0401117},
 primaryClass = {astro-ph},
       adsurl = {https://ui.adsabs.harvard.edu/abs/2004ApJ...612..724Y}
}

@ARTICLE{Nomura2020,
       author = {{Nomura}, Mariko and {Ohsuga}, Ken and {Done}, Chris},
        title = "{Line-driven disc wind in near-Eddington active galactic nuclei: decrease of mass accretion rate due to powerful outflow}",
      journal = {\mnras},
         year = 2020,
        month = may,
       volume = {494},
       number = {3},
        pages = {3616-3626},
          doi = {10.1093/mnras/staa948},
archivePrefix = {arXiv},
       eprint = {1811.01966},
 primaryClass = {astro-ph.HE},
       adsurl = {https://ui.adsabs.harvard.edu/abs/2020MNRAS.494.3616N}
}

@ARTICLE{King2003,
       author = {{King}, A.~R. and {Pounds}, K.~A.},
        title = "{Black hole winds}",
      journal = {\mnras},
         year = 2003,
        month = oct,
       volume = {345},
       number = {2},
        pages = {657-659},
          doi = {10.1046/j.1365-8711.2003.06980.x},
archivePrefix = {arXiv},
       eprint = {astro-ph/0305541},
 primaryClass = {astro-ph},
       adsurl = {https://ui.adsabs.harvard.edu/abs/2003MNRAS.345..657K}
}

@ARTICLE{Nayakshin2001,
       author = {{Nayakshin}, Sergei and {Kallman}, Timothy R.},
        title = "{Accretion Disk Models and Their X-Ray Reflection Signatures. I. Local Spectra}",
      journal = {\apj},
         year = 2001,
        month = jan,
       volume = {546},
       number = {1},
        pages = {406-418},
          doi = {10.1086/318250},
archivePrefix = {arXiv},
       eprint = {astro-ph/0005597},
 primaryClass = {astro-ph},
       adsurl = {https://ui.adsabs.harvard.edu/abs/2001ApJ...546..406N}
}

@ARTICLE{Ross1999,
       author = {{Ross}, R.~R. and {Fabian}, A.~C. and {Young}, A.~J.},
        title = "{X-ray reflection spectra from ionized slabs}",
      journal = {\mnras},
         year = 1999,
        month = jun,
       volume = {306},
       number = {2},
        pages = {461-466},
          doi = {10.1046/j.1365-8711.1999.02528.x},
archivePrefix = {arXiv},
       eprint = {astro-ph/9902325},
 primaryClass = {astro-ph},
       adsurl = {https://ui.adsabs.harvard.edu/abs/1999MNRAS.306..461R}
}

@ARTICLE{Ballantyne2011,
       author = {{Ballantyne}, D.~R. and {McDuffie}, J.~R. and {Rusin}, J.~S.},
        title = "{A Correlation between the Ionization State of the Inner Accretion Disk and the Eddington Ratio of Active Galactic Nuclei}",
      journal = {\apj},
         year = 2011,
        month = jun,
       volume = {734},
       number = {2},
          eid = {112},
        pages = {112},
          doi = {10.1088/0004-637X/734/2/112},
archivePrefix = {arXiv},
       eprint = {1104.1984},
 primaryClass = {astro-ph.HE},
       adsurl = {https://ui.adsabs.harvard.edu/abs/2011ApJ...734..112B}
}

@ARTICLE{Kallman2001,
       author = {{Kallman}, T. and {Bautista}, M.},
        title = "{Photoionization and High-Density Gas}",
      journal = {\apjs},
         year = 2001,
        month = mar,
       volume = {133},
       number = {1},
        pages = {221-253},
          doi = {10.1086/319184},
       adsurl = {https://ui.adsabs.harvard.edu/abs/2001ApJS..133..221K}
}

@ARTICLE{Bianchi2002,
       author = {{Bianchi}, S. and {Matt}, G.},
        title = "{Ionized iron Kalpha lines in AGN X-ray spectra}",
      journal = {\aap},
         year = 2002,
        month = may,
       volume = {387},
        pages = {76-81},
          doi = {10.1051/0004-6361:20020372},
archivePrefix = {arXiv},
       eprint = {astro-ph/0203178},
 primaryClass = {astro-ph},
       adsurl = {https://ui.adsabs.harvard.edu/abs/2002A&A...387...76B}
}

@ARTICLE{Tombesi2010,
       author = {{Tombesi}, F. and {Cappi}, M. and {Reeves}, J.~N. and {Palumbo}, G.~G.~C. and {Yaqoob}, T. and {Braito}, V. and {Dadina}, M.},
        title = "{Evidence for ultra-fast outflows in radio-quiet AGNs. I. Detection and statistical incidence of Fe K-shell absorption lines}",
      journal = {\aap},
         year = 2010,
        month = oct,
       volume = {521},
          eid = {A57},
        pages = {A57},
          doi = {10.1051/0004-6361/200913440},
archivePrefix = {arXiv},
       eprint = {1006.2858},
 primaryClass = {astro-ph.HE},
       adsurl = {https://ui.adsabs.harvard.edu/abs/2010A&A...521A..57T}
}

@ARTICLE{Gofford2015,
       author = {{Gofford}, J. and {Reeves}, J.~N. and {McLaughlin}, D.~E. and {Braito}, V. and {Turner}, T.~J. and {Tombesi}, F. and {Cappi}, M.},
        title = "{The Suzaku view of highly ionized outflows in AGN - II. Location, energetics and scalings with bolometric luminosity}",
      journal = {\mnras},
         year = 2015,
        month = aug,
       volume = {451},
       number = {4},
        pages = {4169-4182},
          doi = {10.1093/mnras/stv1207},
archivePrefix = {arXiv},
       eprint = {1506.00614},
 primaryClass = {astro-ph.HE},
       adsurl = {https://ui.adsabs.harvard.edu/abs/2015MNRAS.451.4169G}
}

@ARTICLE{Gu2025,
       author = {{Gu}, Liyi and {Fukumura}, Keigo and {Kaastra}, Jelle and {Eckart}, Megan and {Ballhausen}, Ralf and {Behar}, Ehud and {Diez}, Camille and {Guainazzi}, Matteo and {Kallman}, Timothy and {Kara}, Erin and {Li}, Chen and {Mehdipour}, Missagh and {Mizumoto}, Misaki and {Ogawa}, Shoji and {Panagiotou}, Christos and {Signorini}, Matilde and {Tanimoto}, Atsushi and {Zhao}, Keqin and {Noda}, Hirofumi and {Miller}, Jon and {Yamada}, Satoshi},
        title = "{Delving into the depths of NGC 3783 with XRISM: III. Birth of an ultrafast outflow during a soft flare}",
      journal = {\aap},
         year = 2025,
        month = dec,
       volume = {704},
          eid = {A146},
        pages = {A146},
          doi = {10.1051/0004-6361/202557189},
archivePrefix = {arXiv},
       eprint = {2512.08448},
 primaryClass = {astro-ph.HE},
       adsurl = {https://ui.adsabs.harvard.edu/abs/2025A&A...704A.146G}
}

@ARTICLE{Narayan1994,
       author = {{Narayan}, Ramesh and {Yi}, Insu},
        title = "{Advection-dominated Accretion: A Self-similar Solution}",
      journal = {\apjl},
         year = 1994,
        month = jun,
       volume = {428},
        pages = {L13},
          doi = {10.1086/187381},
archivePrefix = {arXiv},
       eprint = {astro-ph/9403052},
 primaryClass = {astro-ph},
       adsurl = {https://ui.adsabs.harvard.edu/abs/1994ApJ...428L..13N}
}

@ARTICLE{RamosAlmeida2017,
       author = {{Ramos Almeida}, Cristina and {Ricci}, Claudio},
        title = "{Nuclear obscuration in active galactic nuclei}",
      journal = {Nature Astronomy},
         year = 2017,
        month = oct,
       volume = {1},
        pages = {679-689},
          doi = {10.1038/s41550-017-0232-z},
archivePrefix = {arXiv},
       eprint = {1709.00019},
 primaryClass = {astro-ph.GA},
       adsurl = {https://ui.adsabs.harvard.edu/abs/2017NatAs...1..679R}
}

@ARTICLE{Lodders2009,
       author = {{Lodders}, K. and {Palme}, H. and {Gail}, H.-P.},
        title = "{Abundances of the Elements in the Solar System}",
      journal = {Landolt B{\"o}rnstein},
         year = 2009,
        month = jan,
       volume = {4B},
        pages = {712},
          doi = {10.1007/978-3-540-88055-4_34},
archivePrefix = {arXiv},
       eprint = {0901.1149},
 primaryClass = {astro-ph.EP},
       adsurl = {https://ui.adsabs.harvard.edu/abs/2009LanB...4B..712L}
}

@ARTICLE{Mehdipour2025,
       author = {{Mehdipour}, Missagh and {Kaastra}, Jelle S. and {Eckart}, Megan E. and {Gu}, Liyi and {Ballhausen}, Ralf and {Behar}, Ehud and {Diez}, Camille M. and {Fukumura}, Keigo and {Guainazzi}, Matteo and {Hagino}, Kouichi and et al.},
        title = "{Delving into the depths of NGC 3783 with XRISM: I. Kinematic and ionization structure of the highly ionized outflows}",
      journal = {\aap},
         year = 2025,
        month = jul,
       volume = {699},
          eid = {A228},
        pages = {A228},
          doi = {10.1051/0004-6361/202555623},
archivePrefix = {arXiv},
       eprint = {2506.09395},
 primaryClass = {astro-ph.HE},
       adsurl = {https://ui.adsabs.harvard.edu/abs/2025A&A...699A.228M}
}

@ARTICLE{Bentz2013,
       author = {{Bentz}, Misty C. and {Denney}, Kelly D. and {Grier}, Catherine J. and {Barth}, Aaron J. and {Peterson}, Bradley M. and {Vestergaard}, Marianne and {Bennert}, Vardha N. and {Canalizo}, Gabriela and {De Rosa}, Gisella and {Filippenko}, Alexei V. and {Gates}, Elinor L. and {Greene}, Jenny E. and {Li}, Weidong and {Malkan}, Matthew A. and {Pogge}, Richard W. and {Stern}, Daniel and {Treu}, Tommaso and {Woo}, Jong-Hak},
        title = "{The Low-luminosity End of the Radius-Luminosity Relationship for Active Galactic Nuclei}",
      journal = {\apj},
         year = 2013,
        month = apr,
       volume = {767},
       number = {2},
          eid = {149},
        pages = {149},
          doi = {10.1088/0004-637X/767/2/149},
archivePrefix = {arXiv},
       eprint = {1303.1742},
 primaryClass = {astro-ph.CO},
       adsurl = {https://ui.adsabs.harvard.edu/abs/2013ApJ...767..149B}
}

@ARTICLE{Kaspi2000,
       author = {{Kaspi}, Shai and {Smith}, Paul S. and {Netzer}, Hagai and {Maoz}, Dan and {Jannuzi}, Buell T. and {Giveon}, Uriel},
        title = "{Reverberation Measurements for 17 Quasars and the Size-Mass-Luminosity Relations in Active Galactic Nuclei}",
      journal = {\apj},
         year = 2000,
        month = apr,
       volume = {533},
       number = {2},
        pages = {631-649},
          doi = {10.1086/308704},
archivePrefix = {arXiv},
       eprint = {astro-ph/9911476},
 primaryClass = {astro-ph},
       adsurl = {https://ui.adsabs.harvard.edu/abs/2000ApJ...533..631K}
}

@ARTICLE{Tashiro2025,
       author = {{Tashiro}, Makoto and {Kelley}, Richard and {Watanabe}, Shin and {Maejima}, Hironori and {Reichenthal}, Lillian and {Toda}, Kenichi and {Hartz}, Leslie and {Santovincenzo}, Andrea and {Matsushita}, Kyoko and {Yamaguchi}, Hiroya and {Petre}, Robert and {Williams}, Brian and {Guainazzi}, Matteo and {Costantini}, Elisa and {Takei}, Yoh and {Ishisaki}, Yoshitaka and {Fujimoto}, Ryuichi and {Henegar-Leon}, Joy and {Sneiderman}, Gary and {Tomida}, Hiroshi and {Mori}, Koji and {Nakajima}, Hiroshi and {Terada}, Yukikatsu and {Holland}, Matthew and {Loewenstein}, Michael and {Miller}, Eric and {Sawada}, Makoto and {Kallman}, Timothy and {Kaastra}, Jelle and {Done}, Chris and {Enoto}, Teruaki and {Bamba}, Aya and {Corrales}, Lia and {Ueda}, Yoshihiro and {Kara}, Erin and {Zhuravleva}, Irina and {Fujita}, Yutaka and {Arai}, Yoshitaka and {Audard}, Marc and {Awaki}, Hisamitsu and {Ballhausen}, Ralf and {Baluta}, Chris and {Bando}, Nobutaka and {Behar}, Ehud and {Bialas}, Thomas and {Boissay-Malaquin}, Rozenn and {Brenneman}, Laura and {Brown}, Gregory V. and {Chiao}, Meng and {Cumbee}, Renata and {de Vries}, Cor and {den Herder}, Jan-Willem and {D{\'\i}az Trigo}, Mar{\'\i}a and {DiPirro}, Michael and {Dotani}, Tadayasu and {Carrero}, Jacobo Ebrero and {Ebisawa}, Ken and {Eckart}, Megan and {Eckert}, Dominique and {Eguchi}, Satoshi and {Ezoe}, Yuichiro and {Ferrigno}, Carlo and {Foster}, Adam and {Fukazawa}, Yasushi and {Fukushima}, Kotaro and {Furuzawa}, Akihiro and {Gallo}, Luigi C. and {Garcia Martinez}, Javier and {Gorter}, Nathalie and {Grim}, Martin and {Gu}, Liyi and {Hagino}, Kouichi and {Hamaguchi}, Kenji and {Hatsukade}, Isamu and {Hayashi}, Katsuhiro and {Hayashi}, Takayuki and {Hell}, Natalie and {Hodges-Kluck}, Edmund and {Horiuchi}, Takafumi and {Hornschemeier}, Ann and {Hoshino}, Akio and {Ichinohe}, Yuto and {Ikuta}, Chisato and {Iizuka}, Ryo and {Ishi}, Daiki and {Ishida}, Manabu and {Ishihama}, Naoki and {Ishikawa}, Kumi and {Ishimura}, Kosei and {Jaffe}, Tess and {Katsuda}, Satoru and {Kanemaru}, Yoshiaki and {Kenyon}, Steven and {Kilbourne}, Caroline and {Kimball}, Mark and {Kitamoto}, Shunji and {Kobayashi}, Shogo and {Kohmura}, Takayoshi and {Kubota}, Aya and {Leutenegger}, Maurice A. and {Maeda}, Yoshitomo and {Markevitch}, Maxim and {Matsumoto}, Hironori and {Matsuzaki}, Keiichi and {McCammon}, Dan and {McLaughlin}, Brian and {McNamara}, Brian and {Mernier}, Fran{\c{c}}ois and {Miko}, Joseph and {Miller}, Jon M. and {Minesugi}, Kenji and {Mitani}, Shinji and {Mitsuishi}, Ikuyuki and {Mizumoto}, Misaki and {Mizuno}, Tsunefumi and {Mukai}, Koji and {Murakami}, Hiroshi and {Mushotzky}, Richard and {Nakazawa}, Kazuhiro and {Natsukari}, Chikara and {Ness}, Jan-Uwe and {Nigo}, Kenichiro and {Nishiyama}, Mari and {Nobukawa}, Kumiko and {Nobukawa}, Masayoshi and {Noda}, Hirofumi and {Odaka}, Hirokazu and {Ogawa}, Mina and {Ogawa}, Shoji and {Ogorzalek}, Anna and {Okajima}, Takashi and {Okamoto}, Atsushi and {Ota}, Naomi and {Ozaki}, Masanobu and {Paltani}, Stephane and {Plucinsky}, Paul and {Porter}, F. Scott and {Pottschmidt}, Katja and {Quero}, Jose Antonio and {Sasaki}, Takahiro and {Sato}, Kosuke and {Sato}, Rie and {Sato}, Toshiki and {Sato}, Yoichi and {Seta}, Hiromi and {Shida}, Maki and {Shidatsu}, Megumi and {Shigeto}, Shuhei and {Shipman}, Russel and {Shinozaki}, Keisuke and {Shirron}, Peter and {Simionescu}, Aurora and {Smith}, Randall K. and {Soong}, Yang and {Suzuki}, Hiromasa and {Szymkowiak}, Andrew and {Takahashi}, Hiromitsu and {Takeo}, Mai and {Tamagawa}, Toru and {Tamura}, Keisuke and {Tanaka}, Takaaki and {Tanimoto}, Atsushi and {Terashima}, Yuichi and {Tsuboi}, Yohko and {Tsujimoto}, Masahiro and {Tsunemi}, Hiroshi and {Tsuru}, Takeshi Go and {Uchida}, Hiroyuki and {Uchida}, Nagomi and {Uchida}, Yuusuke and {Uchiyama}, Hideki and {Uno}, Shinichiro and {Vink}, Jacco and {Witthoeft}, Michael and {Wolfs}, Rob and {Yamada}, Satoshi and {Yamada}, Shinya and {Yamaoka}, Kazutaka and {Yamasaki}, Noriko and {Yamauchi}, Makoto and {Yamauchi}, Shigeo and {Yanagase}, Keiichi and {Yaqoob}, Tahir and {Yasuda}, Susumu and {Yoneyama}, Tomokage and {Yoshida}, Tessei and {Yukita}, Mihoko},
        title = "{X-Ray Imaging and Spectroscopy Mission}",
      journal = {\pasj},
         year = 2025,
        month = sep,
       volume = {77},
        pages = {S1-S9},
          doi = {10.1093/pasj/psaf023},
       adsurl = {https://ui.adsabs.harvard.edu/abs/2025PASJ...77S...1T}
}

@ARTICLE{Harrison2013,
       author = {{Harrison}, Fiona A. and {Craig}, William W. and {Christensen}, Finn E. and {Hailey}, Charles J. and {Zhang}, William W. and {Boggs}, Steven E. and {Stern}, Daniel and {Cook}, W. Rick and {Forster}, Karl and {Giommi}, Paolo and {Grefenstette}, Brian W. and {Kim}, Yunjin and {Kitaguchi}, Takao and {Koglin}, Jason E. and {Madsen}, Kristin K. and {Mao}, Peter H. and {Miyasaka}, Hiromasa and {Mori}, Kaya and {Perri}, Matteo and {Pivovaroff}, Michael J. and {Puccetti}, Simonetta and {Rana}, Vikram R. and {Westergaard}, Niels J. and {Willis}, Jason and {Zoglauer}, Andreas and {An}, Hongjun and {Bachetti}, Matteo and {Barri{\`e}re}, Nicolas M. and {Bellm}, Eric C. and {Bhalerao}, Varun and {Brejnholt}, Nicolai F. and {Fuerst}, Felix and {Liebe}, Carl C. and {Markwardt}, Craig B. and {Nynka}, Melania and {Vogel}, Julia K. and {Walton}, Dominic J. and {Wik}, Daniel R. and {Alexander}, David M. and {Cominsky}, Lynn R. and {Hornschemeier}, Ann E. and {Hornstrup}, Allan and {Kaspi}, Victoria M. and {Madejski}, Greg M. and {Matt}, Giorgio and {Molendi}, Silvano and {Smith}, David M. and {Tomsick}, John A. and {Ajello}, Marco and {Ballantyne}, David R. and {Balokovi{\'c}}, Mislav and {Barret}, Didier and {Bauer}, Franz E. and {Blandford}, Roger D. and {Brandt}, W. Niel and {Brenneman}, Laura W. and {Chiang}, James and {Chakrabarty}, Deepto and {Chenevez}, Jerome and {Comastri}, Andrea and {Dufour}, Francois and {Elvis}, Martin and {Fabian}, Andrew C. and {Farrah}, Duncan and {Fryer}, Chris L. and {Gotthelf}, Eric V. and {Grindlay}, Jonathan E. and {Helfand}, David J. and {Krivonos}, Roman and {Meier}, David L. and {Miller}, Jon M. and {Natalucci}, Lorenzo and {Ogle}, Patrick and {Ofek}, Eran O. and {Ptak}, Andrew and {Reynolds}, Stephen P. and {Rigby}, Jane R. and {Tagliaferri}, Gianpiero and {Thorsett}, Stephen E. and {Treister}, Ezequiel and {Urry}, C. Megan},
        title = "{The Nuclear Spectroscopic Telescope Array (NuSTAR) High-energy X-Ray Mission}",
      journal = {\apj},
         year = 2013,
        month = jun,
       volume = {770},
       number = {2},
          eid = {103},
        pages = {103},
          doi = {10.1088/0004-637X/770/2/103},
archivePrefix = {arXiv},
       eprint = {1301.7307},
 primaryClass = {astro-ph.IM},
       adsurl = {https://ui.adsabs.harvard.edu/abs/2013ApJ...770..103H}
}

@ARTICLE{Trakhtenbrot2017,
       author = {{Trakhtenbrot}, Benny and {Ricci}, Claudio and {Koss}, Michael J. and {Schawinski}, Kevin and {Mushotzky}, Richard and {Ueda}, Yoshihiro and {Veilleux}, Sylvain and {Lamperti}, Isabella and {Oh}, Kyuseok and {Treister}, Ezequiel and {Stern}, Daniel and {Harrison}, Fiona and {Balokovi{\'c}}, Mislav and {Gehrels}, Neil},
        title = "{BAT AGN Spectroscopic Survey (BASS) - VI. The {\ensuremath{\Gamma}}$_{X}$-L/L$_{Edd}$ relation}",
      journal = {\mnras},
         year = 2017,
        month = sep,
       volume = {470},
       number = {1},
        pages = {800-814},
          doi = {10.1093/mnras/stx1117},
archivePrefix = {arXiv},
       eprint = {1705.01550},
 primaryClass = {astro-ph.GA},
       adsurl = {https://ui.adsabs.harvard.edu/abs/2017MNRAS.470..800T}
}

@ARTICLE{Elitzur2009,
       author = {{Elitzur}, Moshe and {Ho}, Luis C.},
        title = "{On the Disappearance of the Broad-Line Region in Low-Luminosity Active Galactic Nuclei}",
      journal = {\apjl},
         year = 2009,
        month = aug,
       volume = {701},
       number = {2},
        pages = {L91-L94},
          doi = {10.1088/0004-637X/701/2/L91},
archivePrefix = {arXiv},
       eprint = {0907.3752},
 primaryClass = {astro-ph.CO},
       adsurl = {https://ui.adsabs.harvard.edu/abs/2009ApJ...701L..91E}
}

@ARTICLE{Blumenthal1972,
       author = {{Blumenthal}, George R. and {Drake}, G.~W.~F. and {Tucker}, Wallace H.},
        title = "{Ratio of Line Intensities in Helium-Like Ions as a Density Indicator.}",
      journal = {\apj},
         year = 1972,
        month = feb,
       volume = {172},
        pages = {205},
          doi = {10.1086/151340},
       adsurl = {https://ui.adsabs.harvard.edu/abs/1972ApJ...172..205B}
}

@ARTICLE{Tomaru2019,
       author = {{Tomaru}, Ryota and {Done}, Chris and {Ohsuga}, Ken and {Nomura}, Mariko and {Takahashi}, Tadayuki},
        title = "{The thermal-radiative wind in low-mass X-ray binary H1743-322: radiation hydrodynamic simulations}",
      journal = {\mnras},
         year = 2019,
        month = dec,
       volume = {490},
       number = {3},
        pages = {3098-3111},
          doi = {10.1093/mnras/stz2738},
archivePrefix = {arXiv},
       eprint = {1905.11763},
 primaryClass = {astro-ph.HE},
       adsurl = {https://ui.adsabs.harvard.edu/abs/2019MNRAS.490.3098T}
}

@ARTICLE{Waters2021,
       author = {{Waters}, Tim and {Proga}, Daniel and {Dannen}, Randall},
        title = "{Multiphase AGN Winds from X-Ray-irradiated Disk Atmospheres}",
      journal = {\apj},
         year = 2021,
        month = jun,
       volume = {914},
       number = {1},
          eid = {62},
        pages = {62},
          doi = {10.3847/1538-4357/abfbe6},
archivePrefix = {arXiv},
       eprint = {2101.09273},
 primaryClass = {astro-ph.GA},
       adsurl = {https://ui.adsabs.harvard.edu/abs/2021ApJ...914...62W}
}

@ARTICLE{Starling2005,
       author = {{Starling}, R.~L.~C. and {Page}, M.~J. and {Branduardi-Raymont}, G. and {Breeveld}, A.~A. and {Soria}, R. and {Wu}, K.},
        title = "{The X-ray spectrum of NGC 7213 and the Seyfert-LINER connection}",
      journal = {\mnras},
         year = 2005,
        month = jan,
       volume = {356},
       number = {2},
        pages = {727-733},
          doi = {10.1111/j.1365-2966.2004.08493.x},
archivePrefix = {arXiv},
       eprint = {astro-ph/0410333},
 primaryClass = {astro-ph},
       adsurl = {https://ui.adsabs.harvard.edu/abs/2005MNRAS.356..727S}
}
\bibliographystyle{aasjournal}

\end{document}